\documentclass[11pt]{article}
\usepackage{graphicx}

\newcommand{\BABARPubYear}    {04}

\newcommand{\BABARConfNumber} {36}
\newcommand{\SLACPubNumber} {10598}
\newcommand{\LANLNumber} {0408064}

\input babarsym

\long\def\inst#1{\par\nobreak\kern 4pt\nobreak
    {\it #1}\par\vskip 10pt plus 3pt minus 3pt}

\def\pstar     {\ensuremath {p^*}\xspace} 
\def\ldata     {\ensuremath {123 \invfb}}

\def\etal {{\it et al.}}

\begin{document}
{\pagestyle{empty}


\begin{flushright}
\babar-CONF-\BABARPubYear/\BABARConfNumber \\
SLAC-PUB-\SLACPubNumber \\
hep-ex/\LANLNumber \\
July 2004 \\
\end{flushright}

\par\vskip 3cm

\begin{center}
{\Large \bf
Search for Strange Pentaquark Production in \boldmath{$e^+e^-$} Annihilations 
at \boldmath{$\sqrt{s}=10.58$} \gev and in \Y4S Decays \\  
}
\end{center}
\bigskip

\begin{center}
\large The \babar\ Collaboration\\
\mbox{ }\\
\today
\end{center}
\bigskip \bigskip

\begin{center}
\large \bf Abstract
\end{center}
We present a preliminary inclusive search for strange pentaquark production in
$e^+e^-$ interactions at a center-of-mass energy of 10.58 GeV using
123 fb$^{-1}$ of data collected with the \babar\ detector.
We look for the states that have been reported previously:
the $\Theta^+$(1540), interpreted as a $udud\bar{s}$ state; and the
$\Xi^{--}$(1860) and $\Xi^{0}$(1860), 
candidate $dsds\bar{u}$ and $uss(u\bar{u}+d\bar{d})$ states, respectively.
In addition we search for other members of the antidecuplet and corresponding 
octet to which these states are thought to belong.
We find no evidence for the production of such states and set
preliminary limits on their production cross sections as functions of
c.m.\ momentum.
The corresponding limits on the $\Theta^+$(1540) and $\Xi^{--}$(1860) rates per
$e^+e^- \rightarrow q\bar{q}$ event are well below the rates measured for
ordinary baryons of similar mass.

\vfill

\begin{center}
Submitted to the 32$^{\rm nd}$ International Conference on 
High-Energy Physics, ICHEP 04,\\
16 August---22 August 2004, Beijing, China
\end{center}

\vspace{1.0cm}
\begin{center}
{\em Stanford Linear Accelerator Center, Stanford University, 
Stanford, CA 94309} \\ \vspace{0.1cm}\hrule\vspace{0.1cm}
Work supported in part by Department of Energy contract DE-AC03-76SF00515.
\end{center}

\newpage
} 

\begin{center}
\small

The \babar\ Collaboration,
\bigskip

%
B.~Aubert,
R.~Barate,
D.~Boutigny,
F.~Couderc,
J.-M.~Gaillard,
A.~Hicheur,
Y.~Karyotakis,
J.~P.~Lees,
V.~Tisserand,
A.~Zghiche
\inst{Laboratoire de Physique des Particules, F-74941 Annecy-le-Vieux, France }
% ...

\end{center}\newpage

\section{Introduction} 
\label{sec:Introduction} 
 
Recently several experimental groups have reported observations of a
new, manifestly exotic (B=1, S=1) baryon resonance, 
called the $\Theta^+$(1540)~\cite{prl91012003}-\cite{lasttheta}, with
an unusually narrow width ($\Gamma<$8 \mevcc) for a particle in
this mass range that has open channels to strong decay.  
Also, the NA49 experiment reported evidence for an additional 
narrow exotic (B=1, Q$=$S$=-$2) state, called $\Xi^{--}$, 
as well as corresponding $\Xi^0$ state, with masses
close to 1862 \mevcc~\cite{prl92:042003}.
More recently the H1 collaboration reported a narrow ($\Gamma<30$ \mevcc) 
exotic charmed (B=1, C$=-$1) resonance, $\Thc$, with a mass of 3099 \mevcc.
The simplest quark assignments consistent with the quantum numbers of 
$\Theta^+$, $\Xi^{--}$ and $\Thc$ are ($udud\bar{s}$), ($dsds\bar{u}$) and
($udud\bar{c}$), respectively;
therefore these observed states are regarded as pentaquark candidates.

\begin{figure}[hbt] 
\begin{center} 
\scalebox{0.7}{ 
\includegraphics{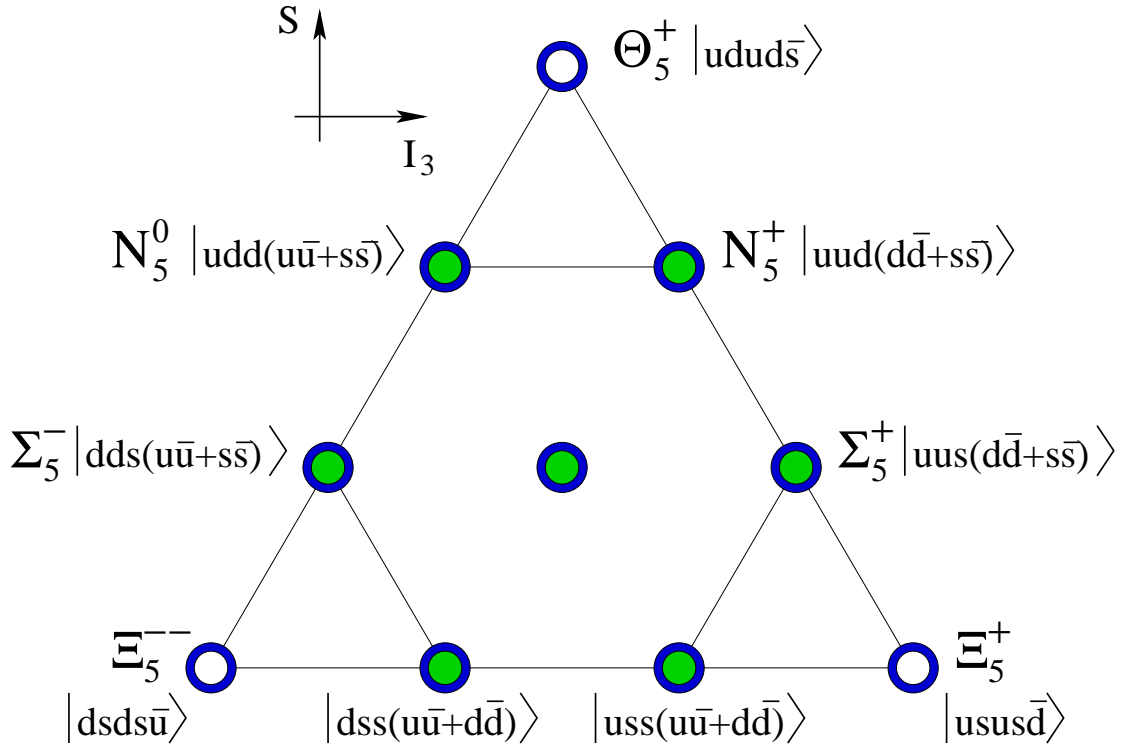}} 
\end{center} 
\vspace{-0.6truecm}
\caption {Quark structure of the antidecuplet (annuli) and octet (filled 
circles) that are generally assumed for the lowest-mass pentaquarks.
Strangeness increases in the vertical direction and the third
component of isospin in the horizontal.}
\label{fig:anti-decouplet} 
\end{figure}

These results have prompted a surge of pentaquark searches in experimental data
of many kinds, mostly with negative results~\cite{review}.
Several theoretical models~\cite{zp:a359:305,Marek,Jaffe} have been proposed to
describe possible pentaquark structure.
They predict that the lowest mass states containing $u$, $d$ and $s$ quarks
should occupy a spin-1/2 antidecuplet and octet, as
illustrated in Fig.~\ref{fig:anti-decouplet}.
Predictions for the masses of the unobserved $N$ and $\Sigma$ states vary; 
$M_N$ might be anywhere between $\sim$150 \mevcc below $M_\Theta$ and
$\sim M_\Xi$,
and $M_\Sigma$ anywhere between $\sim M_\Theta$ and $\sim$150 \mevcc
above $M_\Xi$.
In order to distinguish pentaquarks from ordinary baryons, 
we adopt a notation that has names corresponding 
to ordinary baryons with similar $s$ quark content, plus a subscript $5$. 
For example, the $\xmpq$ pentaquark ($dss(u\bar{u}+d\bar{d})$) corresponds to
the ordinary $\xm$ ($dss$) baryon.

Although experiments with a baryon in the beam or the target might have some 
advantage in pentaquark production, $\epem$ interactions are also
known for democratic production of hadrons.
Baryons with nonzero beauty, charm, strangeness (up to three units)
and/or orbital
angular momentum have been observed with production rates that appear to
depend on the mass and spin, but not quark content.
If pentaquarks are produced similarly, then
one might expect a pentaquark rate as high as that for an ordinary
baryon of the same mass and spin, i.e.\ about $8\times10^{-4}$ $\Th$ and about
$4\times10^{-5}$ $\xmm$ per \eeHad event 
at $\sqrt{s}= 10.58$ \gev~\cite{ref:pdg2002}.
Decays of $B$ mesons are also known to produce a rather high rate of
baryons and so provide another fertile hunting ground.
In both cases we expect any set of pentaquark states forming a
multiplet of approximately equal mass to be produced at roughly equal rates,
so that access to all narrow states is available.

Here we describe a program of inclusive searches for the reported states
and also the other members of the hypothetical
antidecuplet and octet in data from the \babar\ experiment,
which include both $\epem\to q\bar{q}$, $q=udsc$, events and the
production of \Y4S\ mesons, which decay into $B\bar{B}$ pairs.
We have concentrated on decay modes involving strange particles and
protons, which are easily reconstructed in our detector.
All states except the $N_5$ have nonzero strangeness, and the $N_5$
and $\Sigma_5$ have hidden strangeness divided in some way between the
octet and antidecuplet states.
Since the production mechanism, and hence the momentum spectra,
for pentaquarks is unknown, we consider differential production cross sections 
$d\sigma / d\pstar$ per unit momentum \pstar in the \epem center-of-mass
(c.m.) frame, which are
to first order independent of any model, and applicable to the sum of all 
production processes at $\sqrt{s}=10.58$ \gev. 
After describing some common aspects of the analyses in section 2, we
describe searches in three classes of decay modes in the next three
sections.
Upper limits on pentaquark production are discussed in section 6, and
we summarize the results in section 7.

\section{Data and Simulation}
\label{sec:dataandmc}

\noindent 
The data sample used in the analysis comprises the runs of the \babar\ 
experiment through 2003, 
and amounts to an integrated luminosity of \ldata, roughly 90\% 
(10\%) of which was recorded on (below) the \Y4S resonance at a c.m.\ energy of
$\sqrt{s}=10.58$ (10.54) \gev.
The \babar\ detector is described in detail in Ref.~\cite{ref:babar}.
Here we use charged tracks reconstructed in the 5-layer silicon vertex detector
(SVT) and the 40-layer drift chamber (DCH), 
and identified using a combination of energy loss measured in these two 
subdetectors and Cherenkov angles measured in the detector of internally 
reflected Cherenkov radiation (DIRC).
The charged particle momentum resolution is given by
$(\delta p_T /p_T)^2 = (0.0013p_T)^2 + 0.0045^2$, where $p_T$ is the
momentum transverse to the beam axis in \gevc.
Loose particle identification is required on most particles, giving in
each case nearly 100\% efficiency with a sizable background reduction.
Tighter requirements are made where noted, giving identification efficiencies 
ranging from 80--99\% and background rejection factors of 12--100.
Photons are detected by a CsI(Tl) electromagnetic calorimeter (EMC)
with an energy resolution of 
$\sigma(E)/E=0.023\cdot (E/GeV)^{-1/4}\oplus 0.019$.
Our invariant mass resolution for a given decay mode is better than that of any
experiment reporting the observation of a pentaquark candidate with that
mode available.

Minimal hadronic event selection is performed, as we wish to be as
inclusive as possible and maintain maximum signal efficiency.
We simply require three reconstructed charged tracks in the event.
The vast majority of the events are hadronic; 
lepton-pair events are expected to contribute negligible background in the 
signal regions;
two-photon events and events with a hard initial-state photon contribute 
only at low momenta.
The acceptance for a pentaquark signal from any of these sources is well 
understood; 
if a signal is found, we can attempt to isolate the source with cuts on the 
event properties.

Samples of simulated $e^+e^- \rightarrow q\bar{q}$, where $q=udsc$, are 
generated using the JETSET~\cite{jetset} Monte Carlo generator combined with a 
detailed simulation of the \babar\ detector,
in order to study backgrounds and ``control" particles -- known 
resonances with similar masses and decay modes to the pentaquark in 
question.
By comparing with the same particles reconstructed in the data, we verify
the simulation of the invariant mass resolutions and biases of the detector, 
an essential ingredient in the evaluation of any new signal, or limit 
where no signal is seen.

In order to evaluate pentaquark reconstruction efficiencies,
we simulate pentaquark signals within our standard software with 
special parameter settings of the JETSET generator.
For each pentaquark state we take an existing baryon as a stand-in, 
change its mass and width to match the desired pentaquark properties, 
and force a particular decay mode, with all other JETSET parameters
left unchanged.
The generator thus gives the stand-in particles a momentum spectrum based
on the pentaquark mass, but the stand-in flavor.
We make no attempt here to conserve flavor in the
production process, as we are only concerned with providing an appropriate
number and distribution of additional particles in the event.
We do not know the production mechanism for such five-quark states;
however, to make the measurement we only need the efficiency as a function 
of the relevant variables, in this case momentum and polar angle in the
laboratory frame, so any smooth 
distribution that covers the entire range of these variables is sufficient.
Effects such as the type, multiplicity and proximity of other particles in the 
event can be studied by using different stand-in baryons;
such effects are found to be smaller than the other systematic uncertainties.
The stand-in particles used to generate our signal samples are
summarized in Table~\ref{tab:signalmc}.
 
\begin{table}[hbt]
\begin{center} 
\caption{
Baryons used in the simulation to generate pentaquark signal samples. 
}
\vspace{0.1in}
\begin{small}
\begin{tabular}{|l@{$\rightarrow$}l|l@{\hspace*{.7cm}}l|c|c|r|} \hline 
 \multicolumn{2}{|c|}{Pentaquark}  & \multicolumn{2}{c|}{Existing}    &
            Mass       & Width     & \multicolumn{1}{c|}{\# of}   \\ 
 \multicolumn{2}{|c|}{Signal mode} & \multicolumn{2}{c|}{Baryon Used} & 
            (\mevcc )  & (\mevcc ) & \multicolumn{1}{c|}{Events}  \\ 
\hline
 \multicolumn{2}{|c|}{   }& \multicolumn{2}{c|}{   }&&  &        \\[-.3cm]
\Th         & $pK^0_S$       &&    $\Delta^{+}$& 1540& 1&120,000 \\[.08cm]
\xmm        & $\xm\pim$      &&   $\Delta^{++}$& 1862& 1&120,000 \\[.08cm]
  \multicolumn{2}{|c|}{ }    && $\Sigma_c^{++}$& 1862& 1& 98,000 \\[.08cm]
\xmn        & $\xm\pip$      &&      $\Xi^{*0}$& 1862& 1& 60,000 \\[.08cm]
\Sign       & $\xm K^+$      &&      $\Xi^{*0}$& 1862& 1&120,000 \\[.08cm]
\xmpq       & $\Lambda^0K^-$ &&      $\Xi^{*-}$& 1862& 1& 29,000 \\[.08cm]
\xmp/$N_5^+$& $\Lambda^0K^+$ &&$\bar{\Xi}^{*+}$& 1862& 1& 29,000 \\[.08cm]
\xmn        & $\Lambda^0\KS$ &&      $\Xi^{*0}$& 1862& 1& 29,000 \\[.08cm]
\xmpq       & $ \Sigma^0K^-$ &&      $\Xi^{*-}$& 1862& 1& 27,000 \\[.08cm]
\xmp/$N_5^+$& $ \Sigma^0K^+$ &&$\bar{\Xi}^{*+}$& 1862& 1& 27,000 \\[.08cm]
\xmn        & $ \Sigma^0\KS$ &&      $\Xi^{*0}$& 1862& 1& 25,000 \\[.05cm]
\hline
\end{tabular}
\end{small}
\label{tab:signalmc}
\end{center}
\end{table}

\large
\clearpage

\section{Search for the \boldmath{$\Th(1540) \to p\KS$} }
\label{sec:thetaplus}

We reconstruct \Th candidates in the $p\KS$ decay mode, where $\KS \to
\pim\pip$, with selection criteria designed for high efficiency and
low bias against any production mechanism.
A sample of \KS candidates is obtained from all pairs of oppositely charged 
tracks passing loose pion identification requirements that pass within
6 mm of each other.
Each pair is required to: have a total momentum vector extrapolating
within 6 mm (32 mm) of the interaction point (IP) in the plane transverse
to (along) the beam direction; 
have a positive flight distance, defined as the projection on the
candidate momentum direction of a vector from its point of 
closest approach to the beam axis to its decay point;
and have an invariant mass within 10 \mevcc of the nominal \KS mass.
We also require the helicity angle 
$\theta_H$ of the candidate, defined as the angle between the \pip and the
reconstructed $\pip\pim$ flight direction in the $\pip\pim$ rest frame, to
satisfy $|\cos\theta_H|<0.8$, eliminating background from $\Lambda^0$ decays 
and photon conversions and reducing combinatoric background.

\begin{figure}[hbt]
\begin{center}
\scalebox{0.6}{
\includegraphics[width=15.0cm]{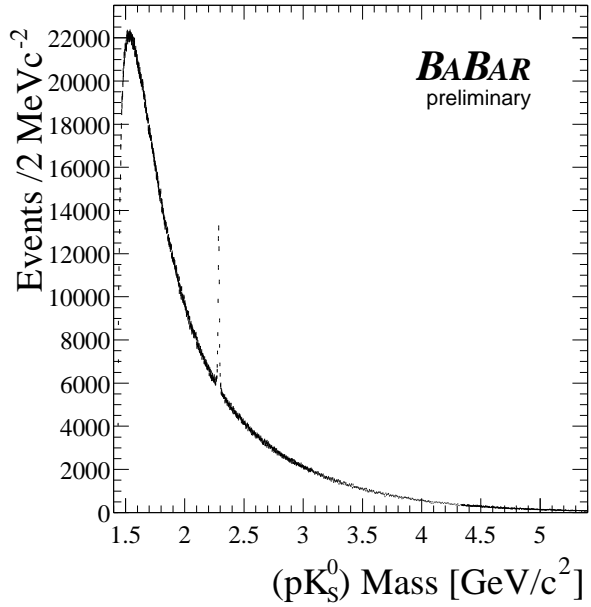}
\includegraphics[width=15.0cm]{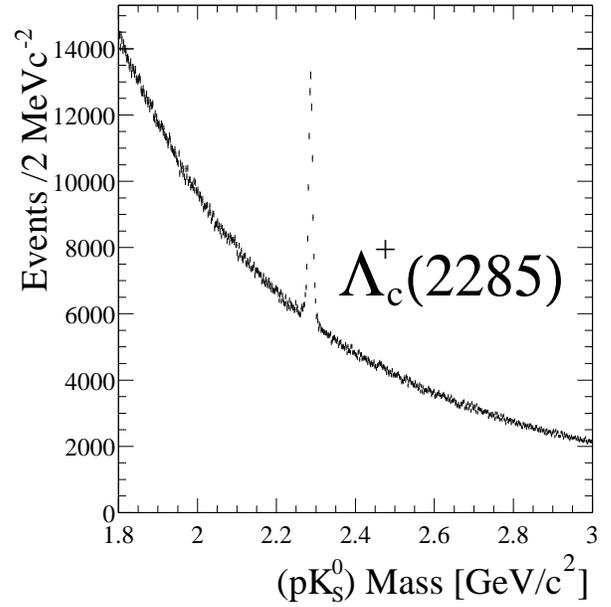}}
\scalebox{0.6}{
\includegraphics[width=15.0cm]{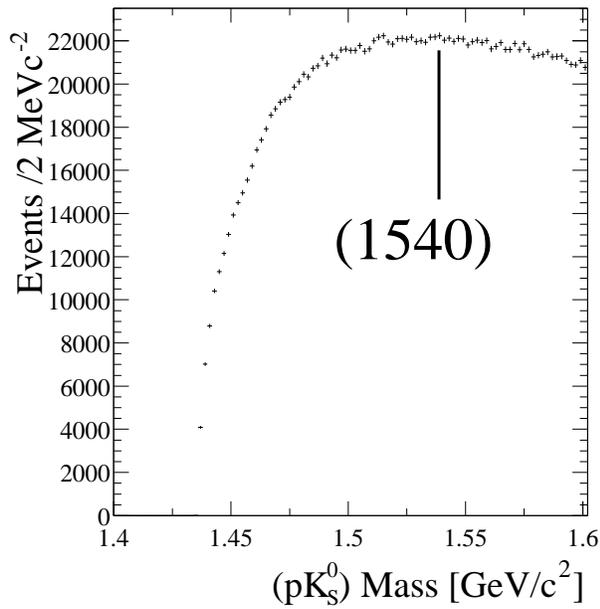}
\includegraphics[width=15.0cm]{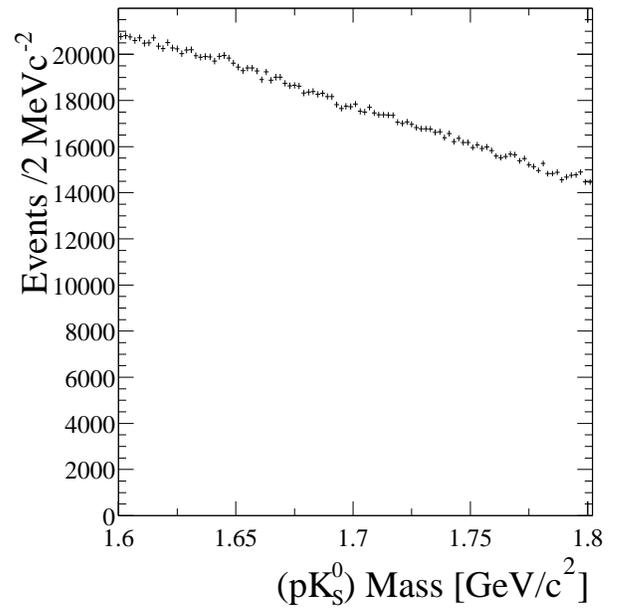}}
 \caption{Distribution of the $p\KS$ invariant mass for combinations
  satisfying all the criteria described in the text.
  The same data are plotted four times in different $p\KS$ mass regions.
}
 \label{fig:thpmass1}
\end{center}
\end{figure}

We combine the surviving \KS candidates with identified $p$ and
$\bar{p}$ tracks that extrapolate within 15 mm (10 cm) of the IP in the plane 
transverse to (along) the beam direction.
The simulated signal reconstruction efficiency varies with \pstar,
the $p\KS$ momentum in the $e^+e^-$ c.m.\ frame, 
from 13\% at low \pstar to 22\% at high \pstar.
The invariant mass distribution of these pairs is
shown in Fig.~\ref{fig:thpmass1}.  
There is a clear peak at 2285 \mevcc from $\Lambda_c^+ \rightarrow p\KS$ 
but no other sharp structure.
The $\Lambda_c^+$ peak (upper right plot) shows a mass resolution of better 
than 6 \mevcc and contains roughly 52,000 entries,
demonstrating our sensitivity to the presence of a narrow resonance.
The lower plots zoom in on the mass ranges 1400--1600 and 1600--1800 \mevcc.
The $\Theta^+_5$ has been reported at values between 1520 and 1540 \mevcc.
There is no enhancement in our data anywhere in this region.
The $p\KS$ decay mode is also possible for the $\Sigma_5^+$ states,
whose masses might be expected to be between that of the \Th and about
2000 \mevcc depending on their strange quark content.
There is no sign of a narrow resonance anywhere in this region.

We consider several additional criteria that might enhance a pentaquark signal.
More stringent requirements on the flight distance of the \KS candidate give a 
much cleaner \KS sample at the expense of efficiency at low momentum.
This enhances the $\Lambda_c^+$ signal to noise, but does not reveal any 
additional structure.
If a pentaquark is produced in an $e^+e^-$ annihilation event, then there
must also be an antibaryon (or anti-pentaquark) in the event, among whose decay
products is either an antiproton or antineutron.
In the case of the $\Theta^+_5$ decaying to $p\KS$, the \KS must have been a 
$K^0$ rather than a $\bar{K}^0$, and there must be a compensating particle in 
the event with strangeness $-$1, which might often be a $K^-$.
We consider all subsets of the $p\KS$ combinations for which there is
also at least one loosely identified $K^-$ and/or $\bar{p}$ track in the event.
In all cases the data quantity is reduced, the $\Lambda_c^+$ signal is
still visible and there is no sign of a pentaquark peak.
Following work by the CLAS and NA49
experiments~\cite{prl92:032001,prl92:042003}, we reconstruct a possible
$N^*(2400)\rightarrow K^-\KS p$ using the selected $p\KS$ pairs and an 
additional loosely identified $K^-$ candidate.
We observe no $N^*$ peak in our data.
Considering various $K^-\KS p$ mass ranges, 
the $p\KS$ mass distribution for $K^-\KS p$ masses near 2400 \mevcc 
shows no pentaquark signal, and is indistinguishable from the $p\KS$ mass 
distributions in nearby $K^-\KS p$ mass ranges.
Requiring in addition a recoil antiproton yields similar results.

\clearpage

In each of the above cases we split the data into ten bins of \pstar, 
in order to enhance our sensitivity to any production mechanism that
gives a momentum spectrum different from that of our background.
The bins are 500 \mevcc wide and cover the \pstar range from zero
to 5 \gevcc, the kinematic limit for a particle of mass 1700 \mevcc.
The background is much smaller at higher \pstar, so we are more sensitive to 
mechanisms that produce harder spectra.
In no case is there any sign of a pentaquark signal.

We quantify these null results for the \Th, assuming a mass of 1540 \mevcc.
In order to reduce model dependence, we consider the ten \pstar bins noted
above and fit a signal plus background function to the $p\KS$ invariant mass 
distribution in each bin.
The natural width of the $\Theta^+_5$ has not been measured; 
the best upper limit, $\Gamma < 8$ \mevcc, is larger than our detector 
resolution, 
so we must consider the range of widths up to this value.
We use a P-wave Breit-Wigner lineshape multiplied by a phase-space factor and 
convolved with a resolution function derived from the $\Lambda^+_c$ data and 
simulation. 
The latter is a sum of two Gaussian distributions with a common center and an
overall rms ranging from 2.5 \mevcc at low \pstar to 1.8 \mevcc at high \pstar;
this is narrower than for the $\Lambda^+_c$ due to the proximity of the 
$\Theta^+_5(1540)$ to threshold. 
For the natural width we consider two possiblities, 1 \mevcc and 8 \mevcc,
corresponding to a very narrow state and the upper limit, respectively.
The fits are performed over a wide range, from threshold to 1800 \mevcc, and 
the background function is a seventh-order polynomial times a threshold 
factor.

\begin{figure}[hbt]
\begin{center}
\scalebox{0.6}{
\includegraphics[width=13.7cm]{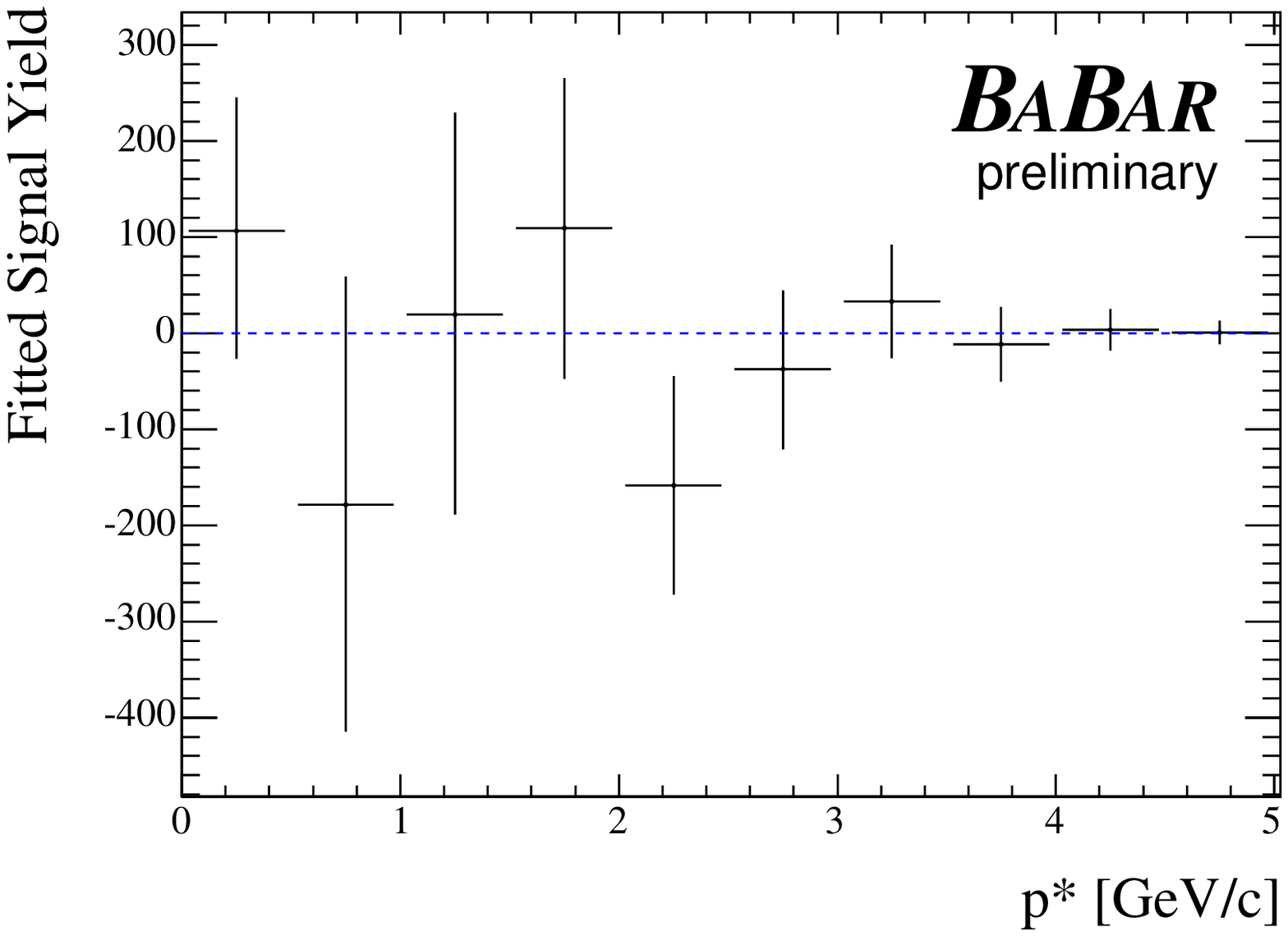}
\includegraphics[width=13.7cm]{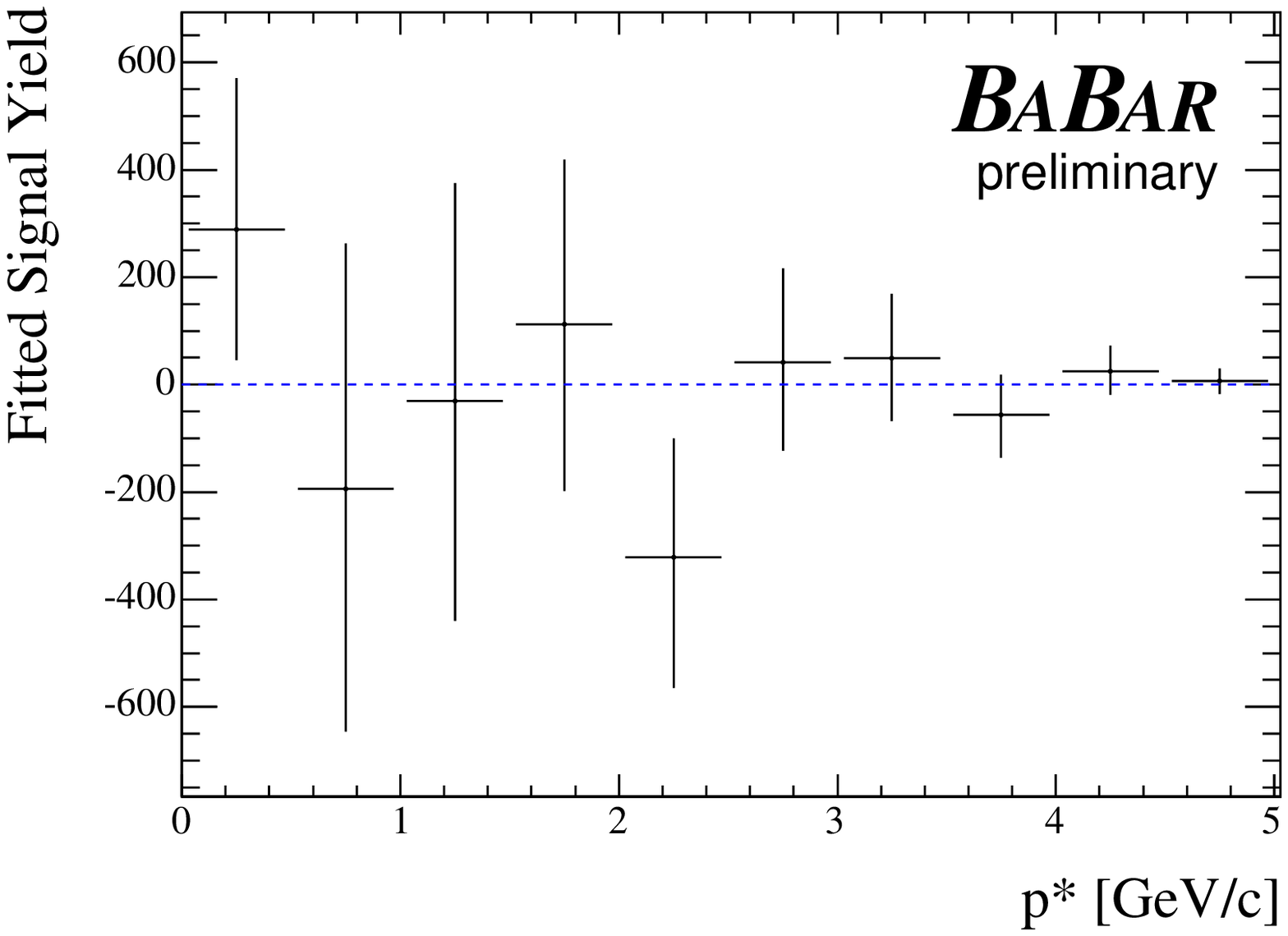}}
\end{center}
\vspace{-0.8truecm}
\caption {The \Th signal yields extracted from the fits to the $p\KS$
  invariant mass distributions, assuming a \Th mass of 1540 \mevcc and
  width $\Gamma=1$ \mevcc (left) and  $\Gamma=8$ \mevcc (right).} 
\label{fig:thppq}
\end{figure}

For the case of our nominal selection criteria (Fig.~\ref{fig:thpmass1}), 
we obtain the signal yields shown in Fig.~\ref{fig:thppq}.
In all \pstar bins the fit quality is good and the signal is
consistent with zero.
There is no positive trend in the data, and the roughly symmetric scatter
of the points about zero indicates low
momentum-dependent bias in the background function.
We consider systematic effects in the fitting procedure by varying 
the signal and background functions and fit range; changes in the
fitted signal are negligible compared with the statistical uncertainties.
Other selection criteria yield similar results.
Since the nominal selection gives the
smallest absolute uncertainties after efficiency corrections, we use it
to set upper limits on the production cross section in section 6.
Varying the mass assumed for the \Th has effects consistent with
statistics; in no case do we observe a signal, and the uncertainties
shown in Fig.~\ref{fig:thppq} are typical.

\clearpage

\section{Search for \boldmath{$\xmm\!\rightarrow\!\Xi^-\pi^-$} and
\boldmath{$\xmn \to \Xi^-\pi^+$} 
}

We next search for pentaquark states decaying into a $\Xi^-$ and a
charged pion,
where $\Xi^- \to \Lambda^0\pim$ and $\Lambda^0\to p\pim$,
including the reported $\xmm (1862)$ and $\xmn (1862)$.
We first reconstruct $\Lambda^0\rightarrow p\pim$ candidates from all pairs of 
charged tracks satisfying loose proton and pion requirements.
Efficient and unbiased selection criteria are again applied:  the
tracks must pass within 6 mm of each other; 
the candidate have a positive flight distance from the IP;
and it must have an invariant mass within 10 \mevcc of the nominal 
$\Lambda^0$ mass.
These candidates are combined with an additional negatively charged track 
passing loose pion identification requirements to form $\Xi^-$ candidates.
These candidates are required to form a good vertex, to have a
positive flight distance from the IP, 
and to have an invariant mass within 20 \mevcc of the nominal $\Xi^-$ mass.
Furthermore the flight distance of the $\Lambda^0$ candidate from the 
$\Lambda^0\pim$ vertex must be positive.
Finally, we combine the $\Xi^-$ candidates with an additional charged track 
consistent with coming from the IP and passing 
loose pion identification requirements.
The cosine of the angle between the reconstructed $\xm$ trajectory, 
extrapolated back to the IP, and the additional track is required to be less 
than 0.998.
This last cut is especially important, since the $\xm$ is charged and has
a long lifetime;
if it has a long flight distance, it can produce a reconstructed track
that, if combined with itself, forms a false peak in the invariant
mass distribution.
The simulated signal reconstruction efficiency varies 
from 6.5\% at low \pstar to 12\% at high \pstar.

\begin{figure}[hbt]
\begin{center}
\scalebox{0.6}{
\includegraphics[width=23.0cm]{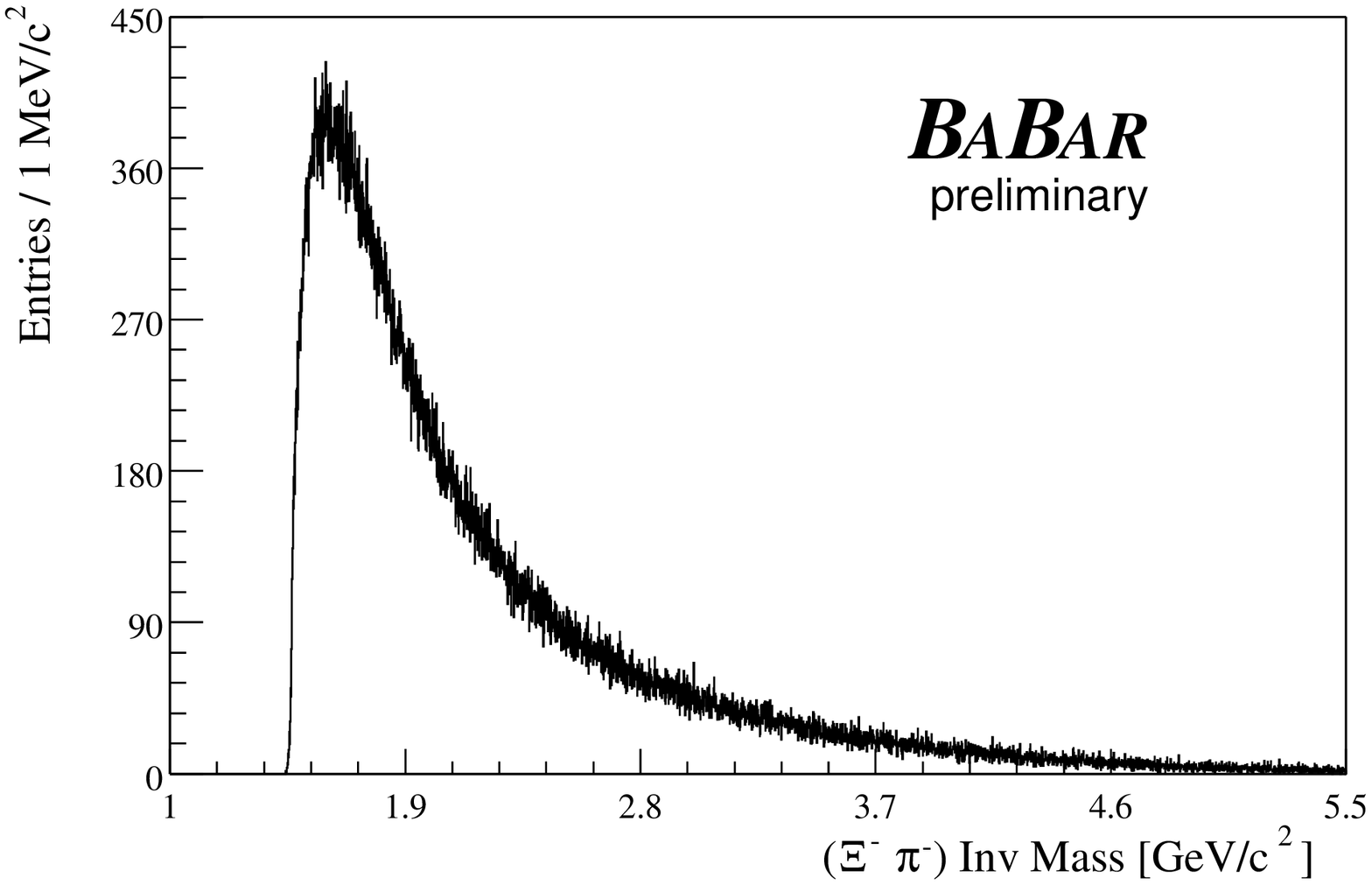}}
\end{center}
\vspace{-0.8truecm}
\caption {$\xm \pim$ invariant mass distribution.} 
\label{fig:mxipim}
\end{figure}

\begin{figure}[hbt]
\begin{center}
\scalebox{0.6}{
\includegraphics[width=23.0cm]{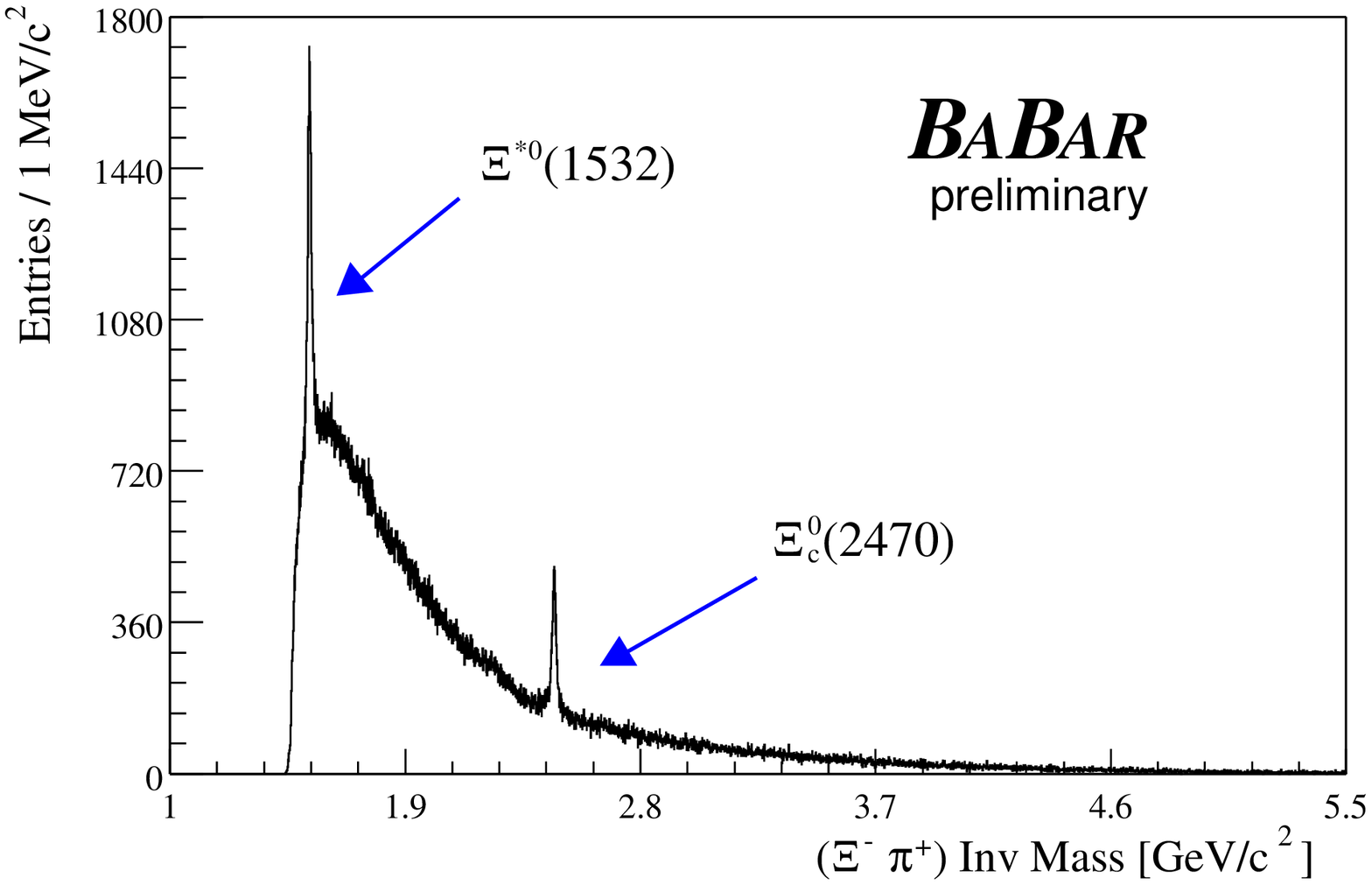}}
\end{center}
\vspace{-0.8truecm}
\caption {$\xm \pip$ invariant mass distribution.} 
\label{fig:mxipip}
\end{figure}

The invariant mass distributions for $\xm \pim$ and $\xm \pip$ 
are shown in Figs.~\ref{fig:mxipim} and \ref{fig:mxipip},
respectively.
In Fig.~\ref{fig:mxipip} there are clear peaks as expected for the
$\Xi^{*0}(1530)$ and $\Xi_c^0(2470)$ baryons, but no other structure
is visible.
In Fig.~\ref{fig:mxipim} 
there is no visible sharp structure at all.

As in the preceding section, we divide the \xmm candidates into ten bins of 
\pstar and find no sign of a pentaquark signal in any bin.
We fit a similar signal plus background function to the $\xm\pim$
invariant mass distribution in each bin.
The resolution function is derived in this case from the $\Xi^{*0}(1530)$ and 
$\Xi_c^0(2470)$ data and simulation, and is described by a Gaussian
with an rms of 8 \mevcc. 
For the Breit-Wigner width we consider two possiblities, 1 and 18 \mevcc, 
corresponding to a very narrow state and the upper limit on the $\xmm$
width, respectively.
The fits are performed over the mass range from 1760 to 1960 \mevcc, and 
the background function is a seventh-order polynomial.

\begin{figure}[hbt]
\begin{center}
\scalebox{0.6}{
\includegraphics[width=21.0cm]{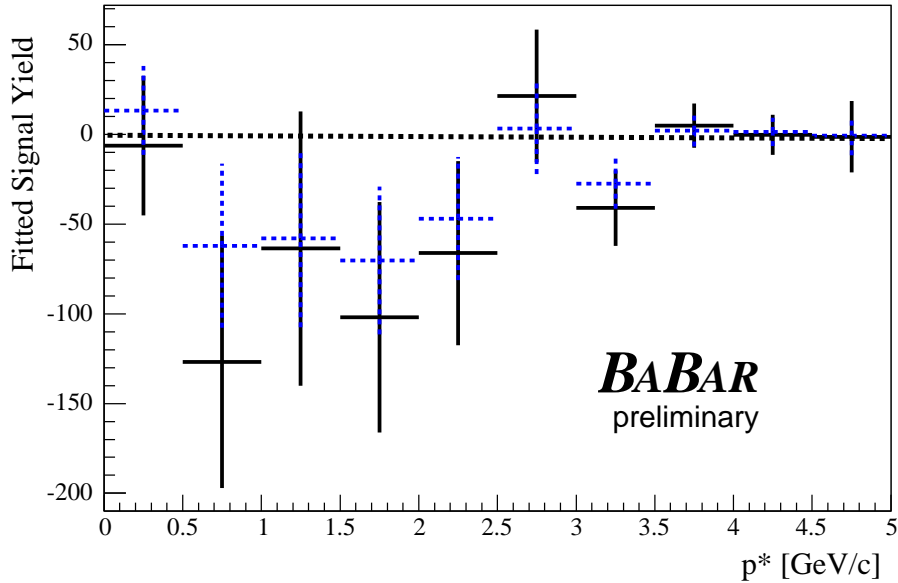}}
\end{center}
\vspace{-0.8truecm}
\caption {The \xmm signal yields extracted from the fits to the $\xm\pim$ 
 invariant mass distributions, assuming a \xmm mass of 1862 \mevcc and 
 width  $\Gamma=1$ (dashed) and  $\Gamma=18$ \mevcc (solid).
} 
\label{fig:xmsignal}
\end{figure}

Fixing the \xmm mass to 1862 \mevcc, we obtain the signal yields shown in 
Fig.~\ref{fig:xmsignal}.
In all \pstar bins the fit quality is good across the full mass range and the 
signal is consistent with zero.
Systematic uncertainties on the fitting procedure are again found to be
negligible compared with the statistical uncertainties, and variations
of the \xmm mass and selection criteria give consistent results.

\section{\boldmath Search for \xmpq, \xmn, \nnz and \nnp in the
\LambK and \SigzK modes}

We search for pentaquark states decaying into a \LambK or \SigzK final state,
where $K$ is a charged or neutral kaon,
$\Sigma^0 \to \Lambda^0\gamma$ and $\Lambda^0\to p\pim$.
These final states give us access to the \xmpq, \xmn, \nnz and \nnp
states (see Fig.~\ref{fig:anti-decouplet}).
The \xmn has been reported with a mass of 1862 \mevcc, and we expect a
quite similar mass for the \xmpq;
since our decay modes include two strange particles, we are sensitive
only to \nnz and \nnp states with substantial $s\bar{s}$ content, and
these might be expected somewhere between 1500 and 1900 \mevcc.

We first reconstruct $\Lambda^0\rightarrow p\pim$ candidates from all pairs of 
charged tracks in which one satisfies tight proton identification
requirements, and the other loose pion identification.
The pairs are required to form a good vertex, to have an angle between
their flight direction (line from the IP to the vertex) and total
momentum at the vertex less than 200 mrad,
and to have an invariant mass within 20 \mevcc of the nominal
$\Lambda^0$ mass.
To reconstruct $\Sigma^0$ candidates, 
The $\Lambda^0$ candidates are combined with neutral (not associated with
any charged track) calorimetric energy deposits of at least 80 \mev that
do not pair with any other neutral deposit to form a $\pi^0$ candidate.
We require a mass difference $M_{p\pi\gamma}-M_{p\pi}$ within 20
\mevcc of the nominal $M_{\Sigma^0}-M_{\Lambda^0}$ value.
Similarly, \KS candidates are reconstructed from pairs of oppositely
charged tracks forming a good vertex, having an angle between
their flight direction and total momentum less than 30 mrad,
and having an invariant mass within 4.5 \mevcc of the nominal \KS
mass.
Charged kaon candidates are required to be consistent with coming from
the IP and to pass tight kaon identification requirements.
The simulated signal reconstruction efficiencies vary from 0.5\% for
the $\Sigma^0 \KS$ mode to 10\% for the $\Lambda^0 K^\pm$ modes at
low \pstar, and increase to 3\% and 25\%, respectively, at high \pstar.

\begin{figure}[t]
\begin{center}
\includegraphics[height=6.0cm]{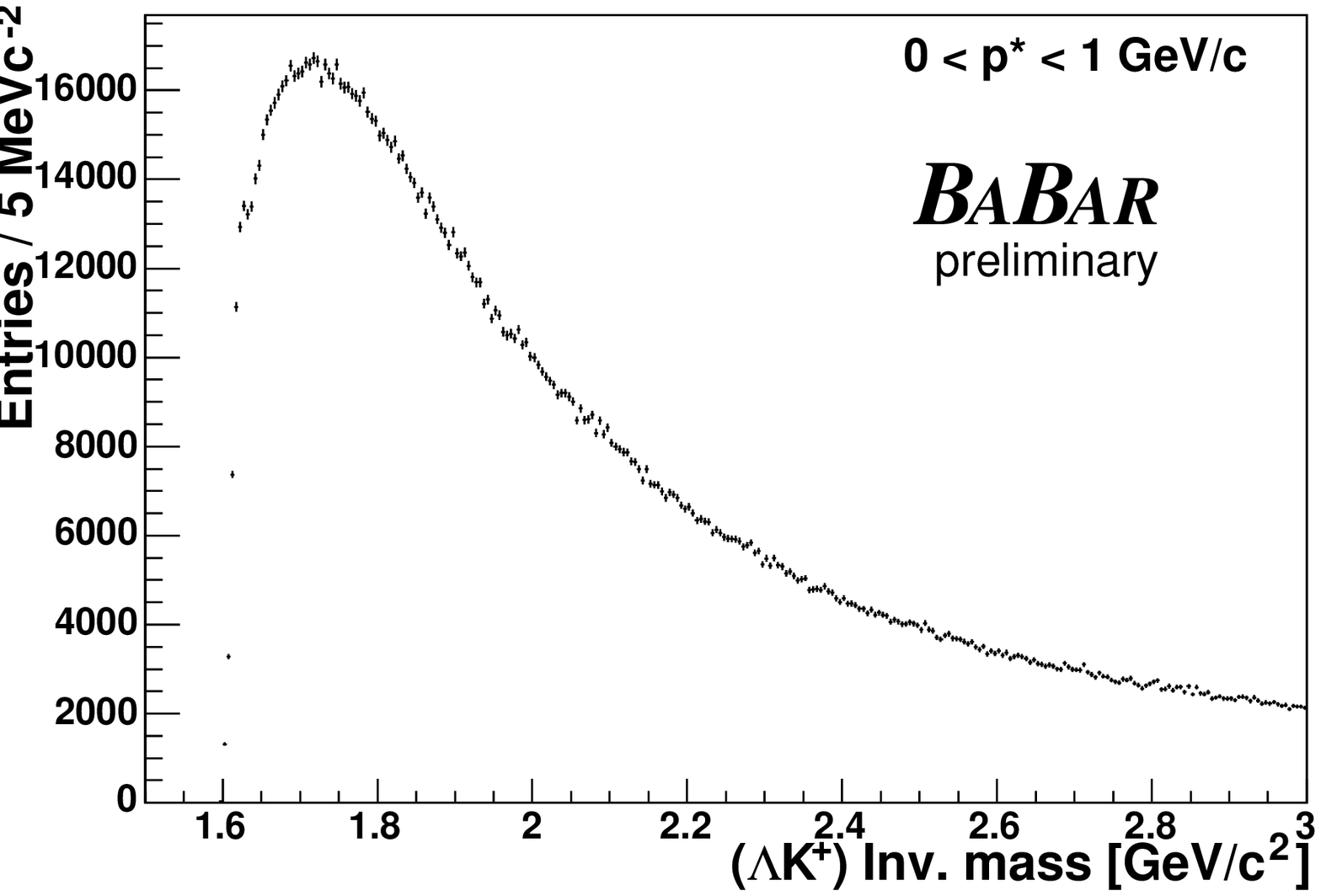}
\includegraphics[height=6.0cm]{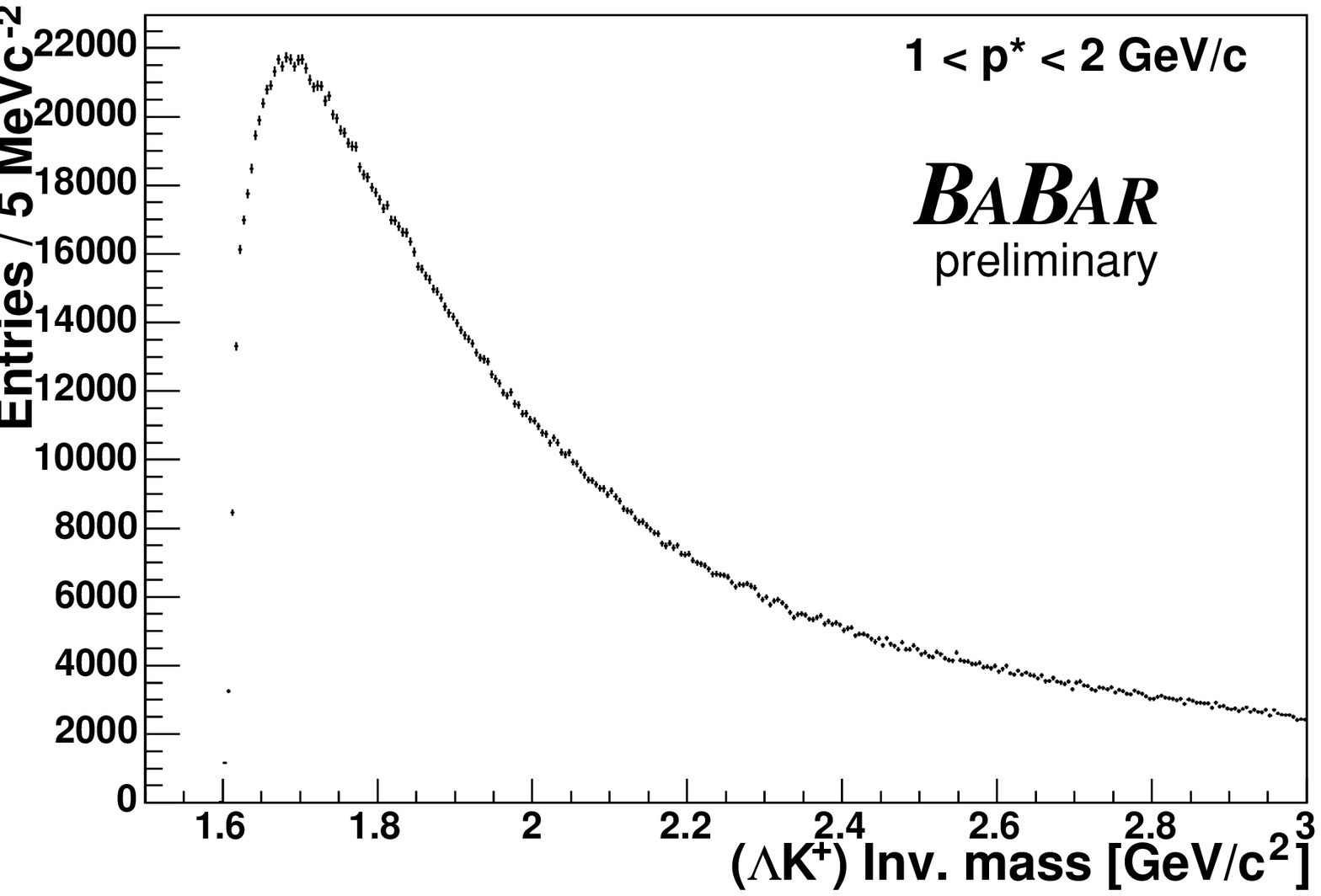}
\includegraphics[height=6.0cm]{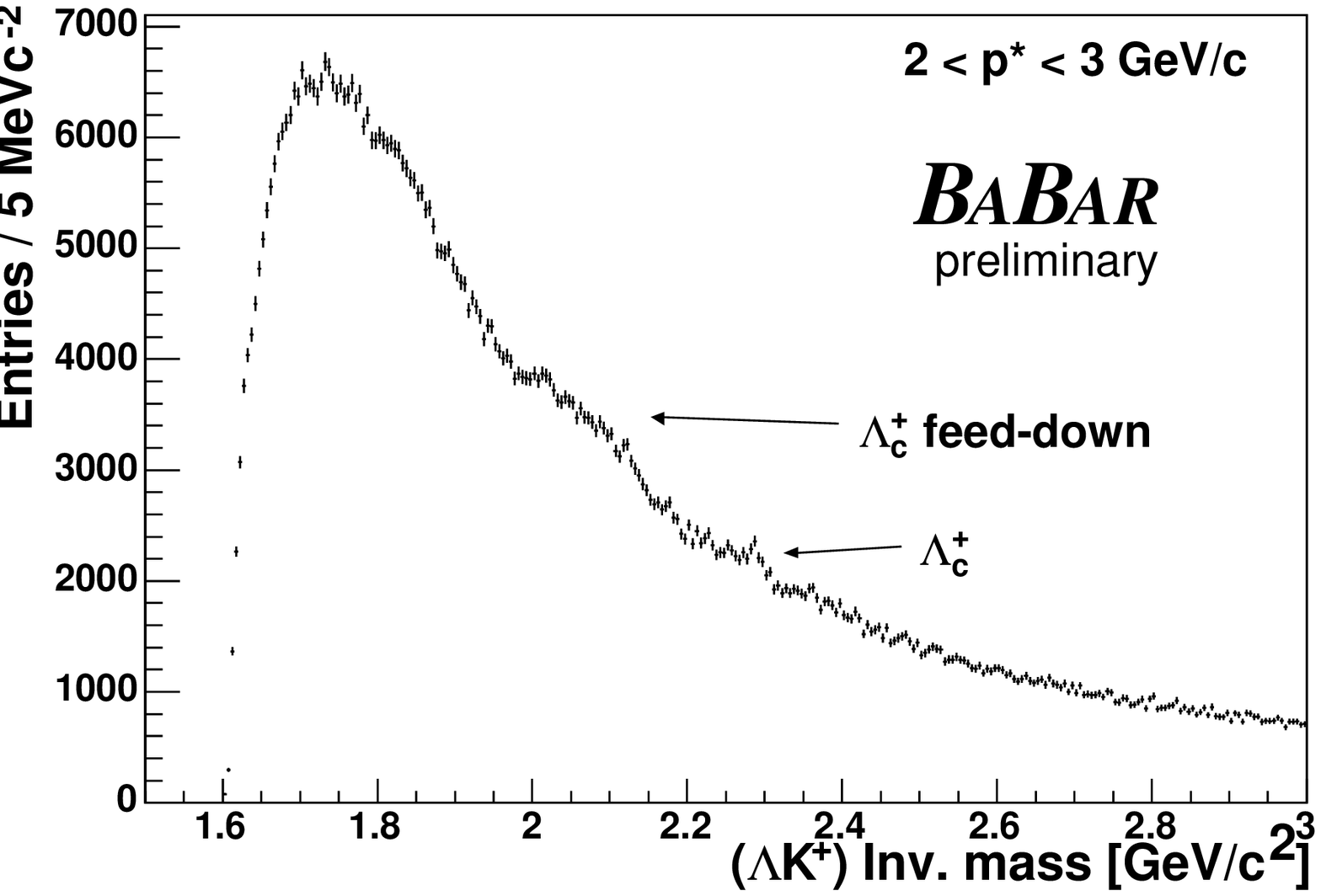}
\includegraphics[height=6.0cm]{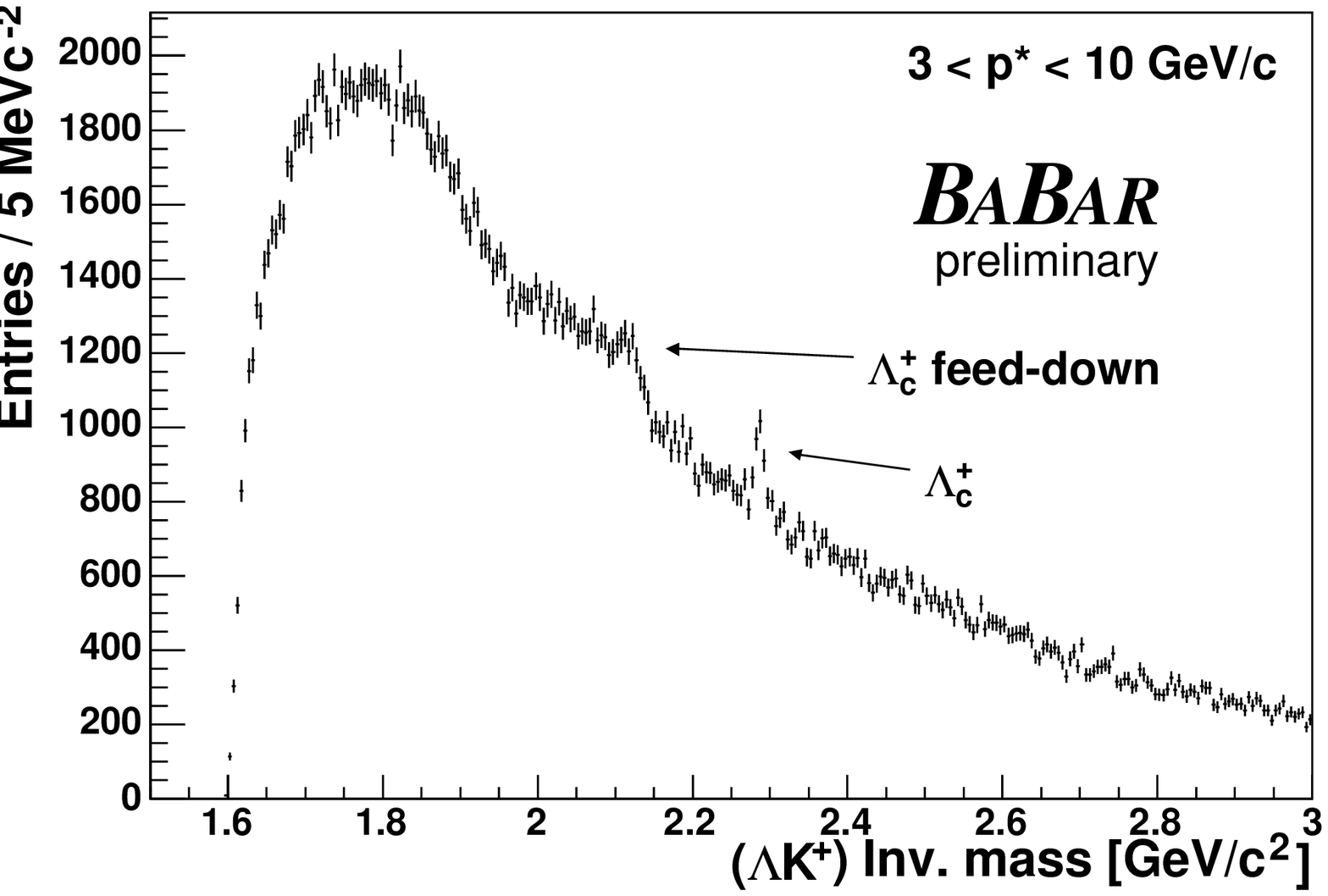}
\vspace{-0.8truecm}
\caption{
Invariant mass distributions of $\LambKp$ for \pstar in four different
ranges.
}
\label{fig:mlamkp}
\end{center}
\end{figure}

\begin{figure}[hbt]
\begin{center}
\vspace*{-0.9truecm}
\includegraphics[height=5.4cm]{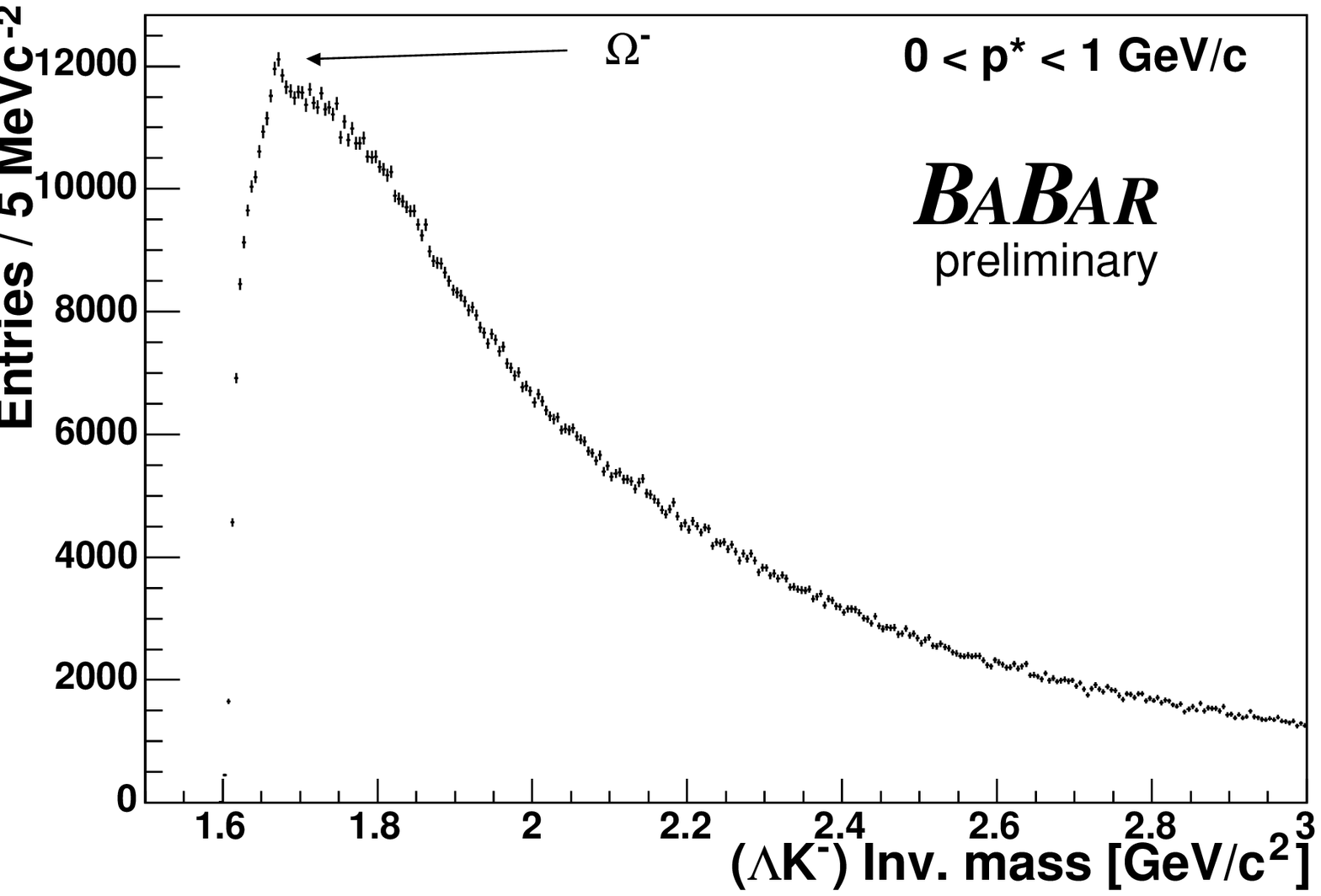}
\includegraphics[height=5.4cm]{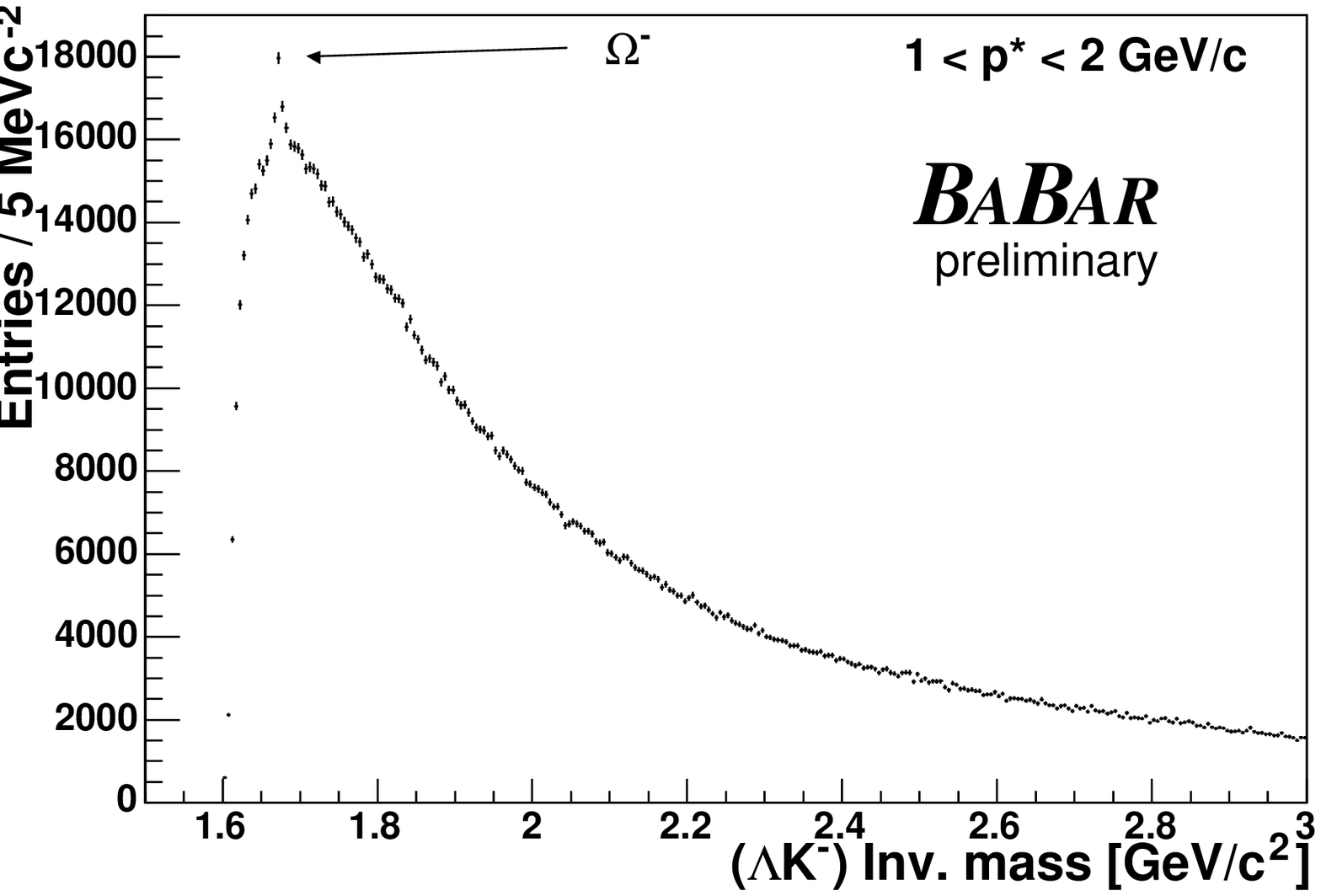}
\vspace*{-0.05truecm}

\includegraphics[height=5.4cm]{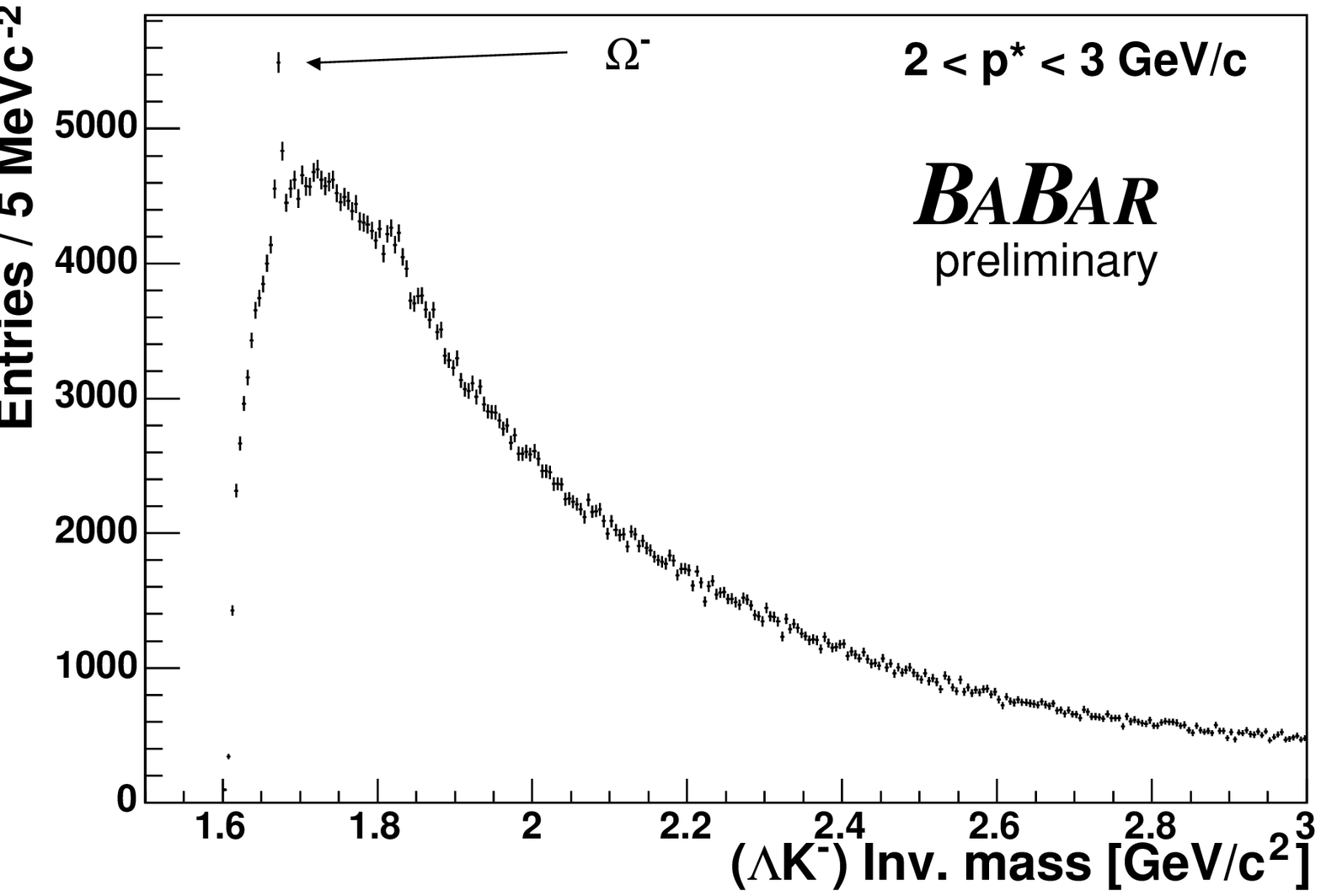}
\includegraphics[height=5.4cm]{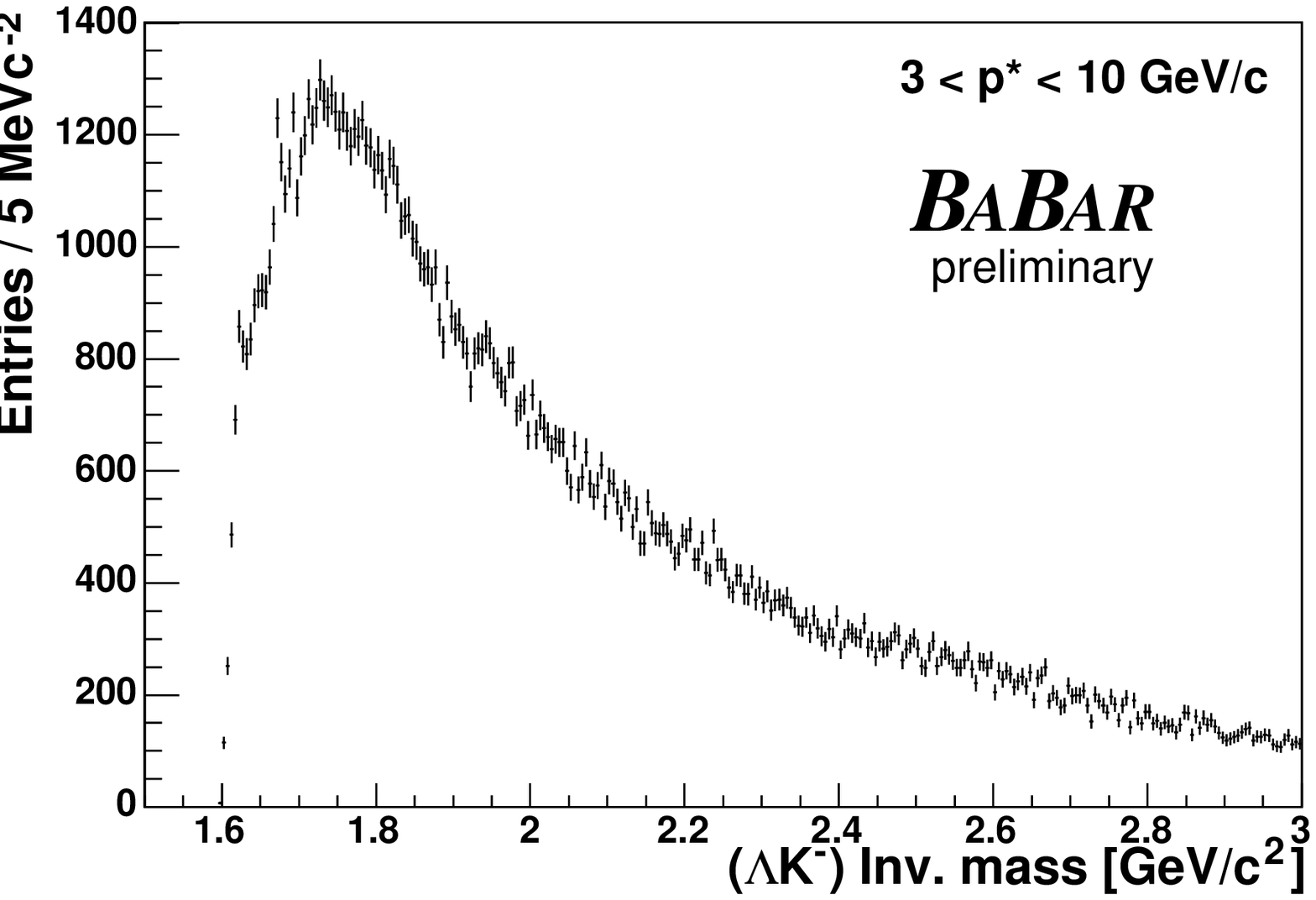}
\vspace{-0.5truecm}
\caption{
Invariant mass distributions of $\LambKm$ for $p^*$ in four different ranges.
}
\label{fig:mlamkm}
\end{center}
\end{figure}

\begin{figure}[hbt]
\begin{center}
\vspace*{-0.5truecm}
\includegraphics[height=5.4cm]{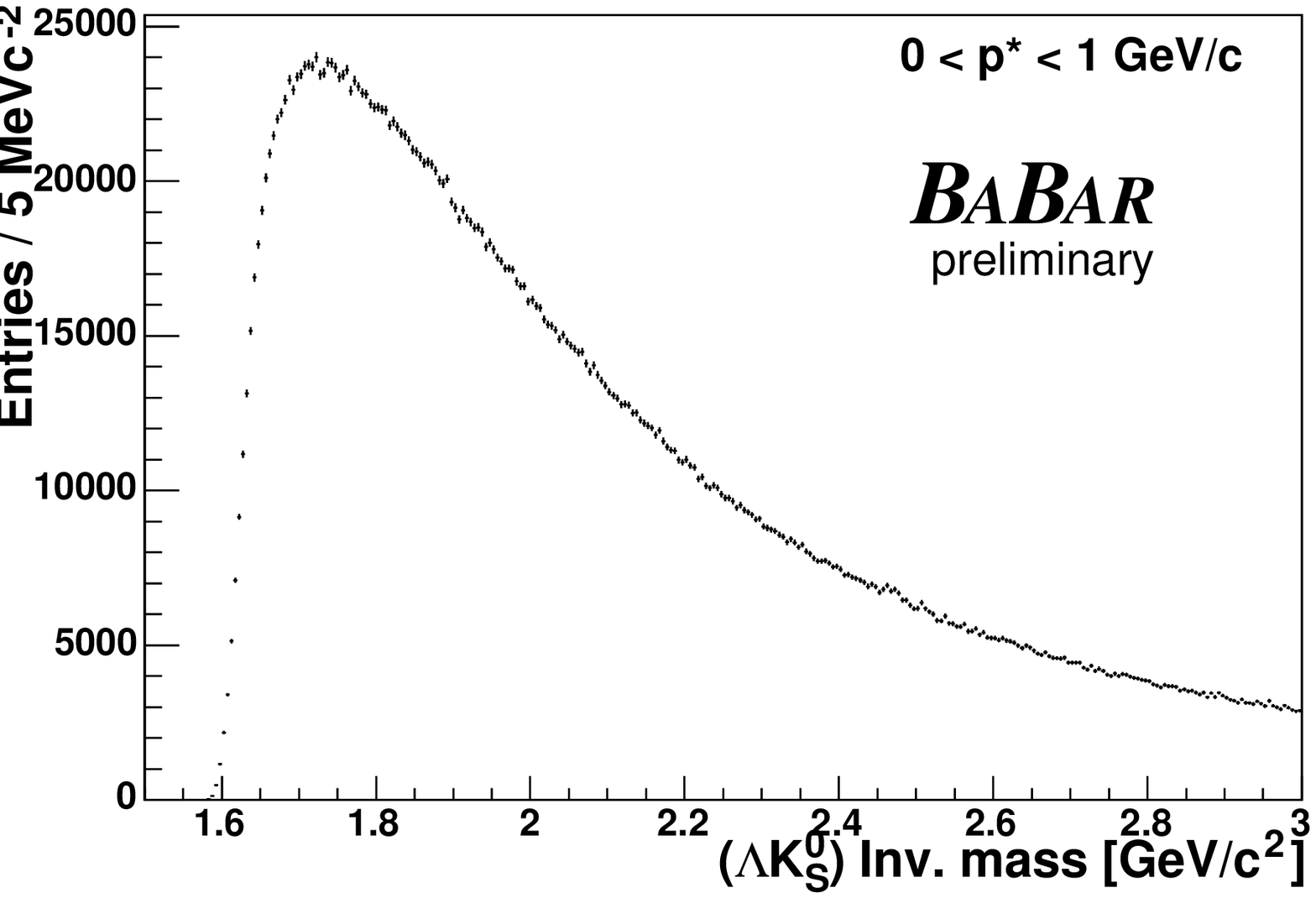}
\includegraphics[height=5.4cm]{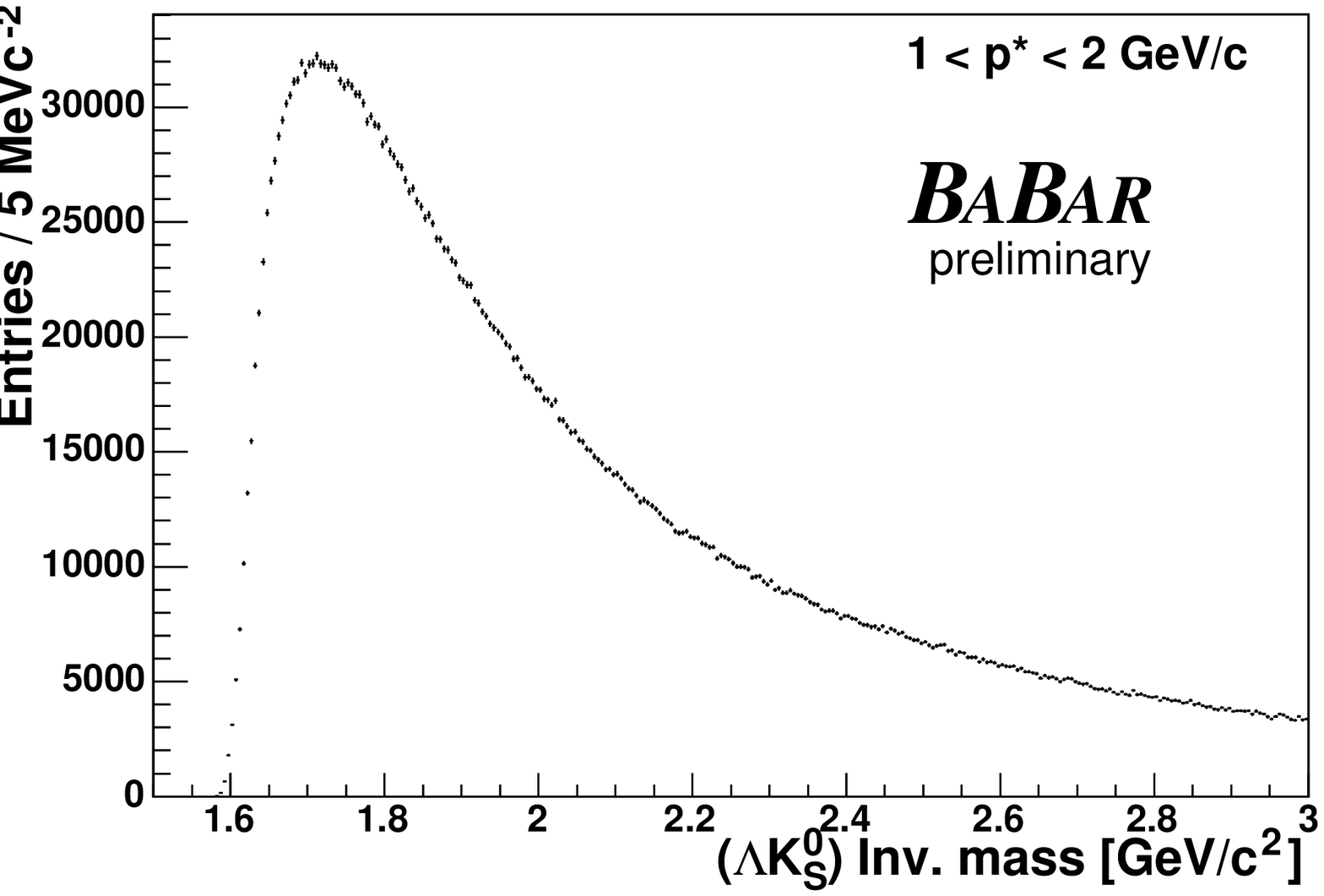}
\vspace*{-0.05truecm}

\includegraphics[height=5.4cm]{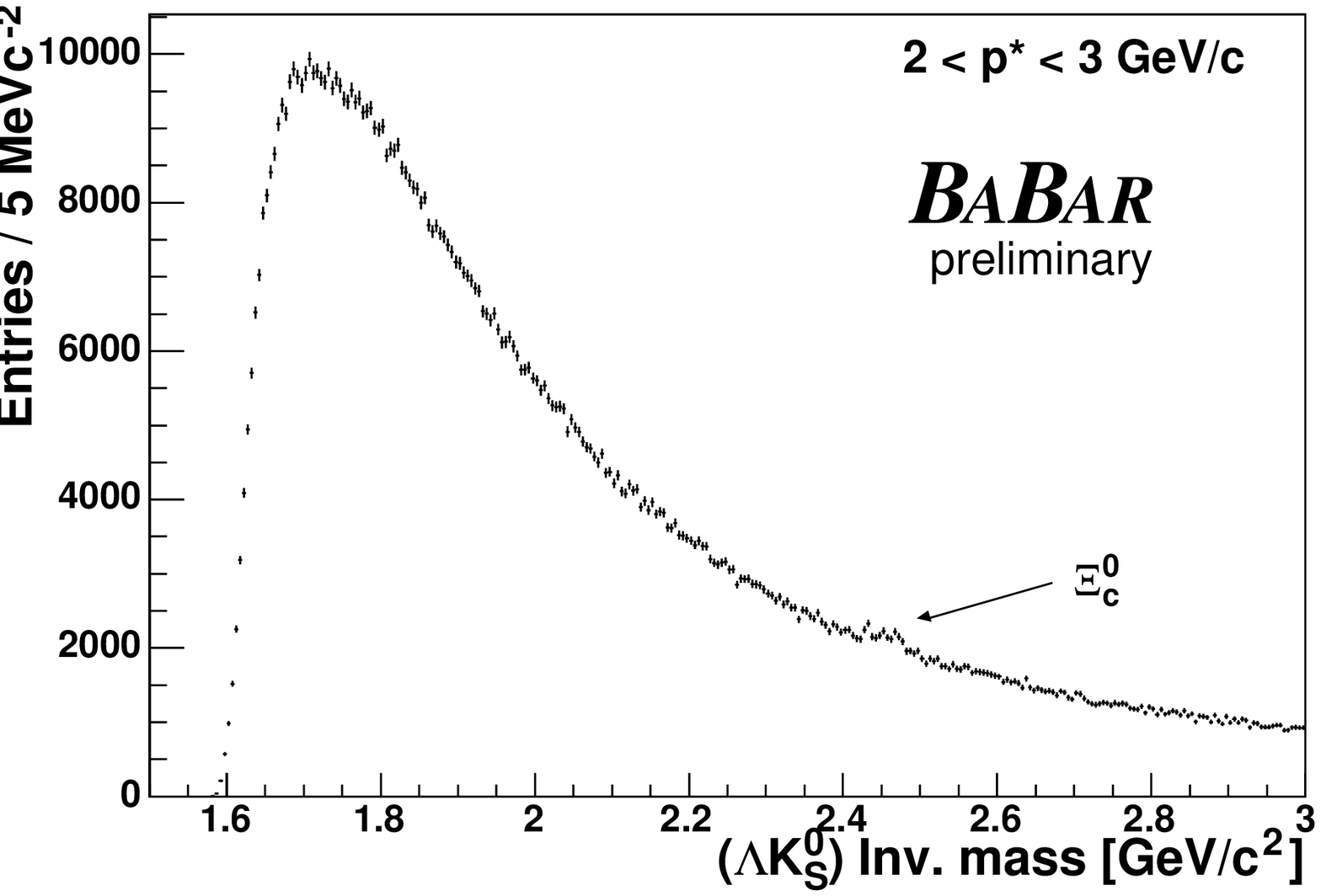}
\includegraphics[height=5.4cm]{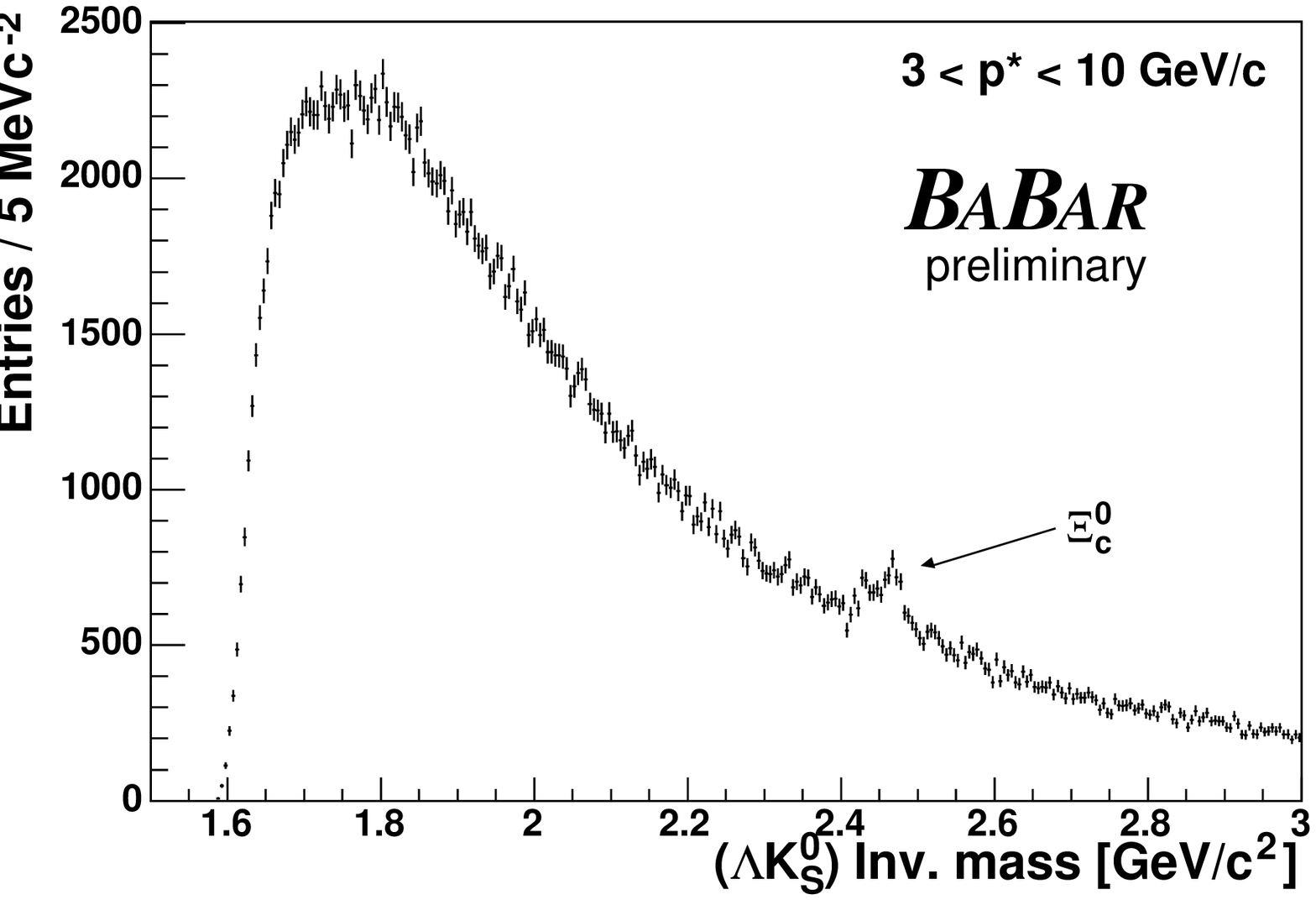}
\vspace{-0.5truecm}
\caption{
Invariant mass distributions of $\LambKs$ for $p^*$ in four different ranges.
}
\label{fig:mlamkz}
\end{center}  
\end{figure}

\begin{figure}[hbt]
\begin{center}
\vspace*{-0.9truecm}
\includegraphics[height=5.4cm]{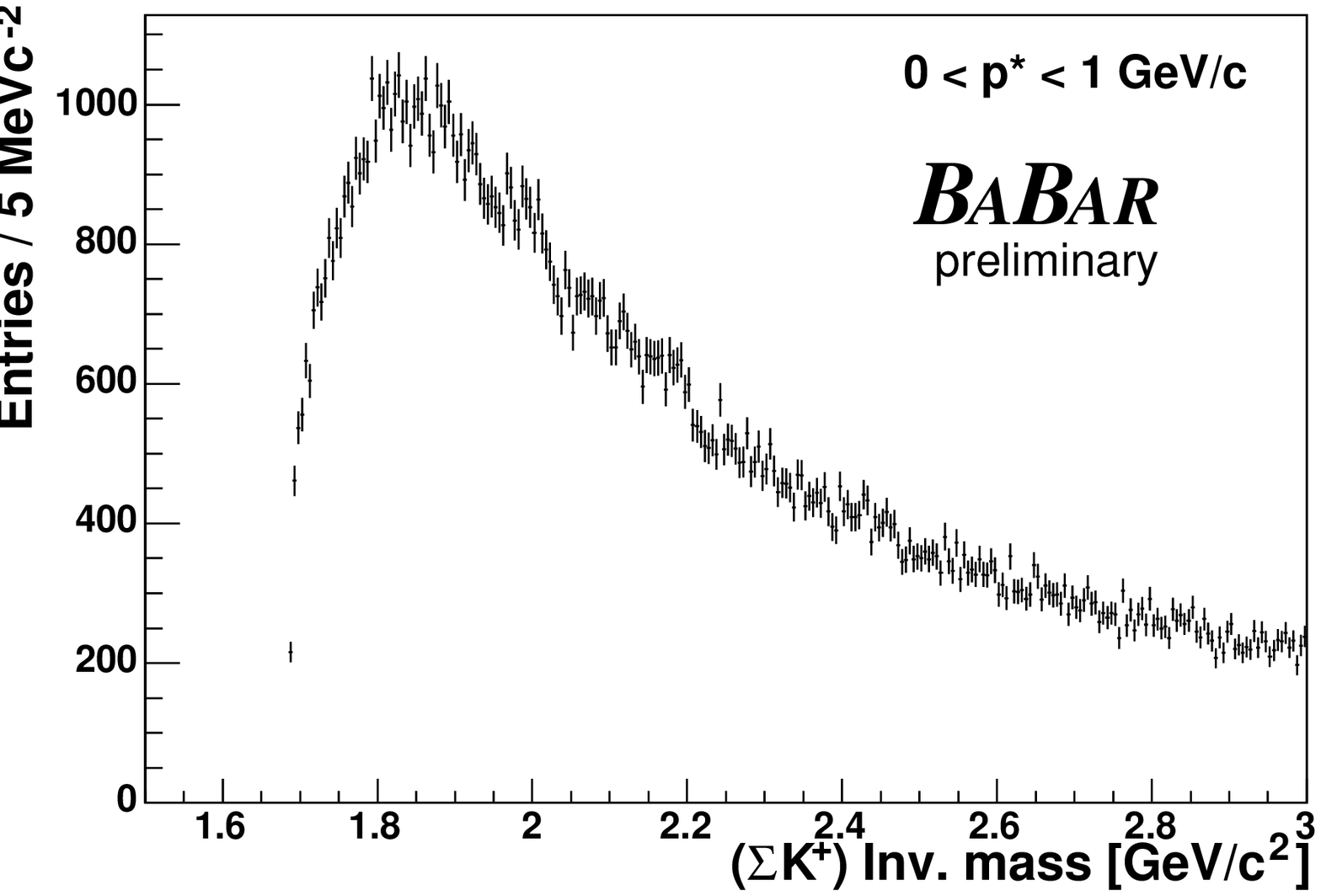}
\includegraphics[height=5.4cm]{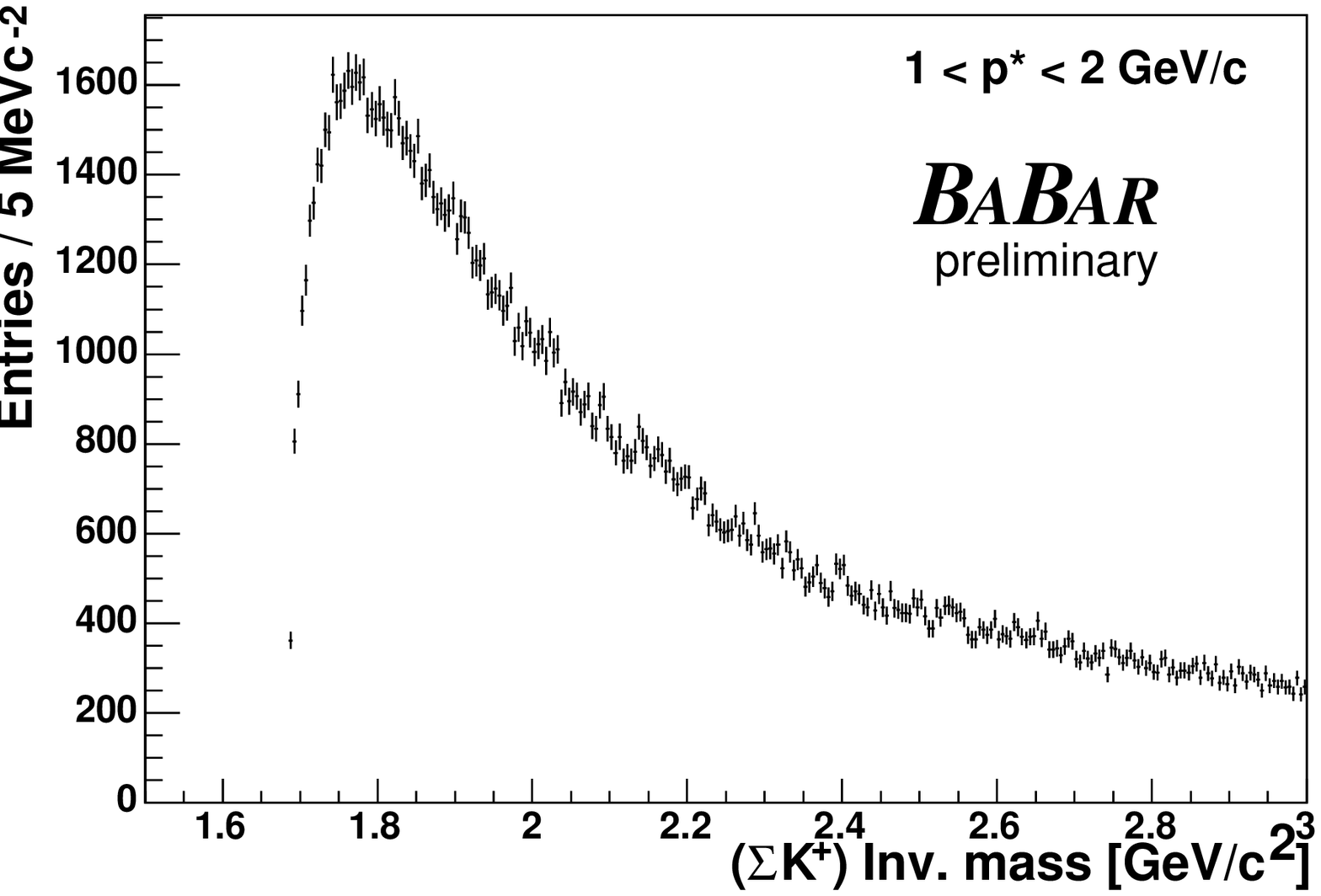}
\vspace*{-0.05truecm}

\includegraphics[height=5.4cm]{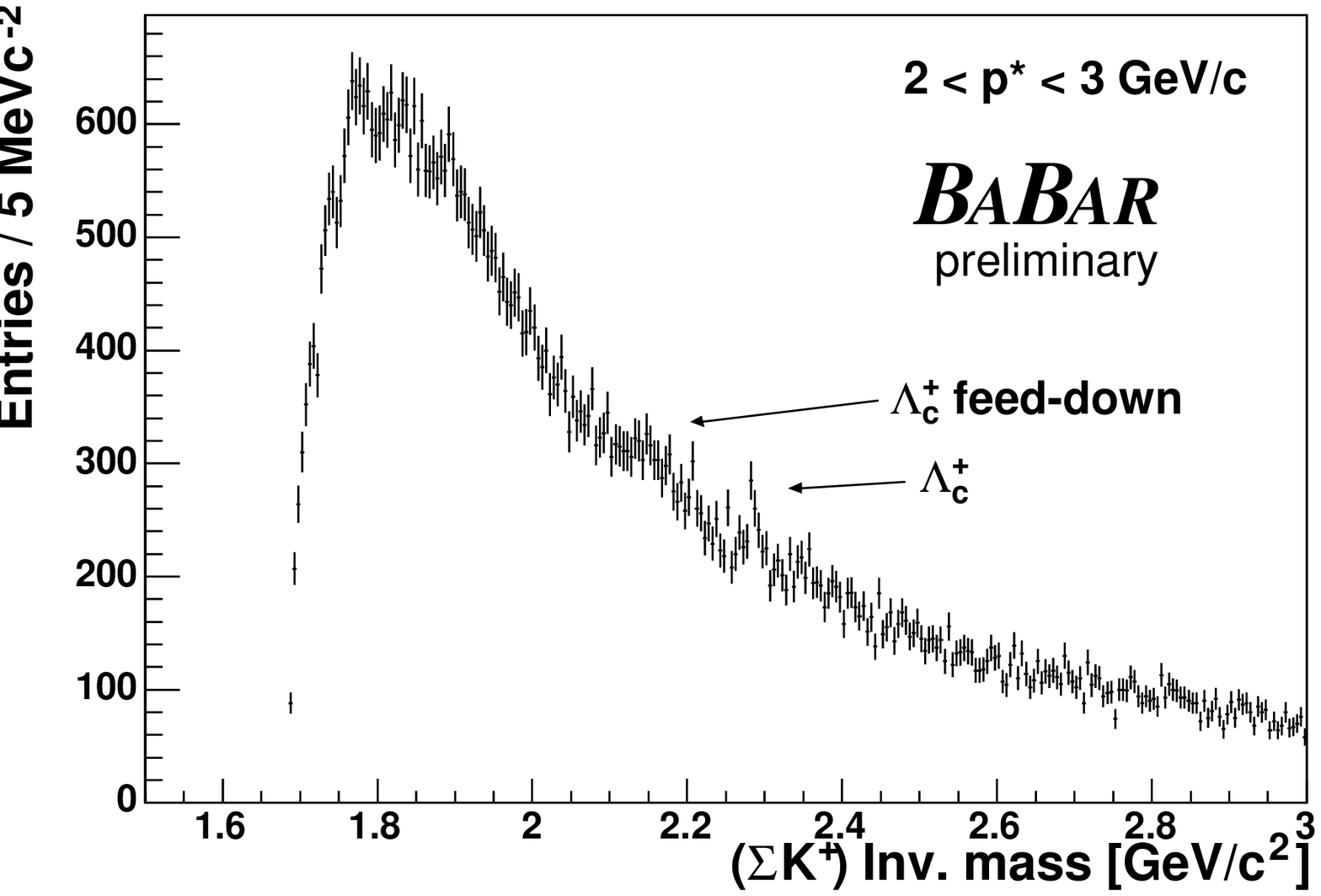}
\includegraphics[height=5.4cm]{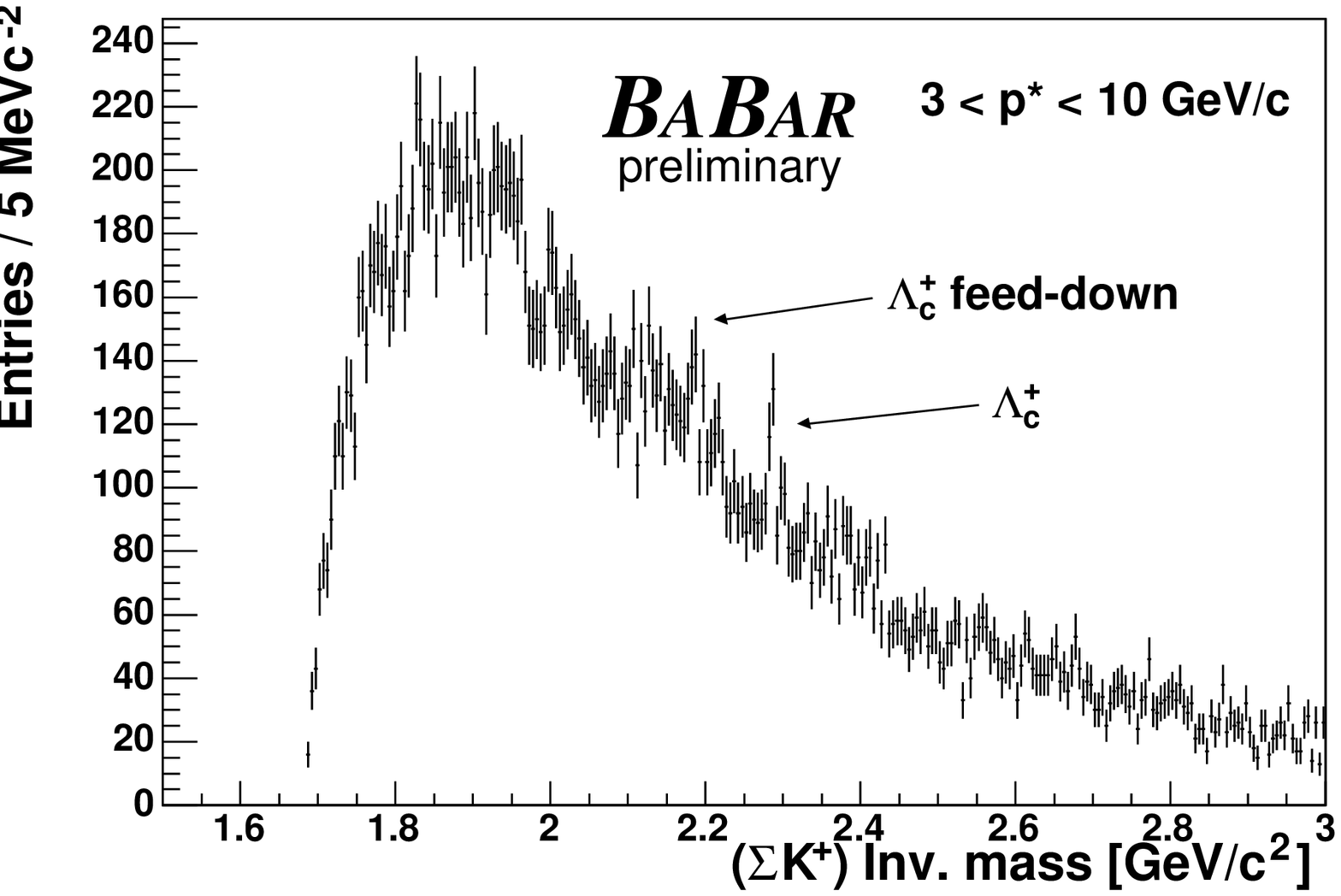}
\vspace{-0.5truecm}
\caption{
Invariant mass distributions of $\SigzKp$ for $p^*$ in four different ranges.}
\label{fig:msigkp}
\end{center}  
\end{figure}

\begin{figure}[hbt]
\begin{center}
\vspace*{-0.5truecm}
\includegraphics[height=5.4cm]{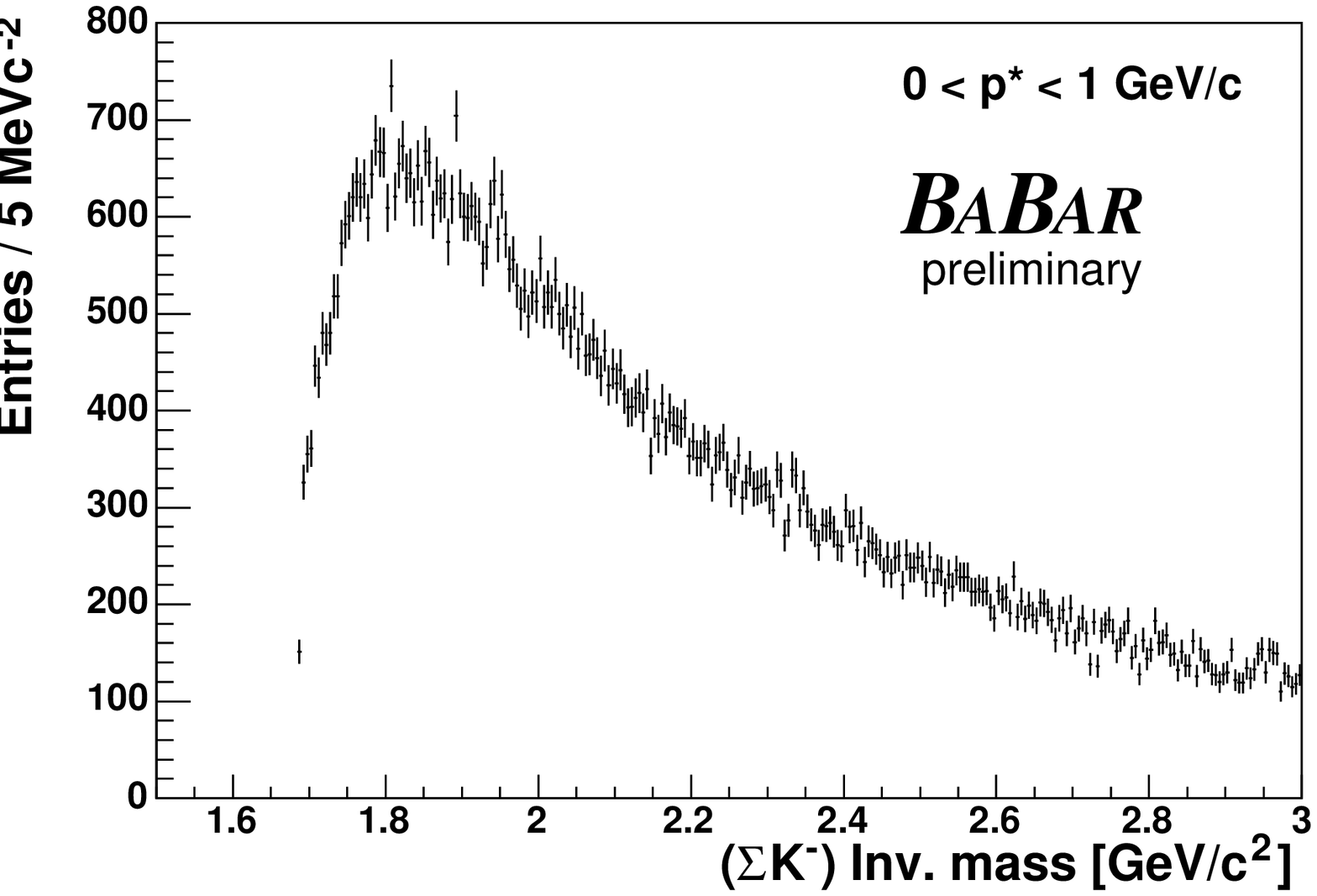}
\includegraphics[height=5.4cm]{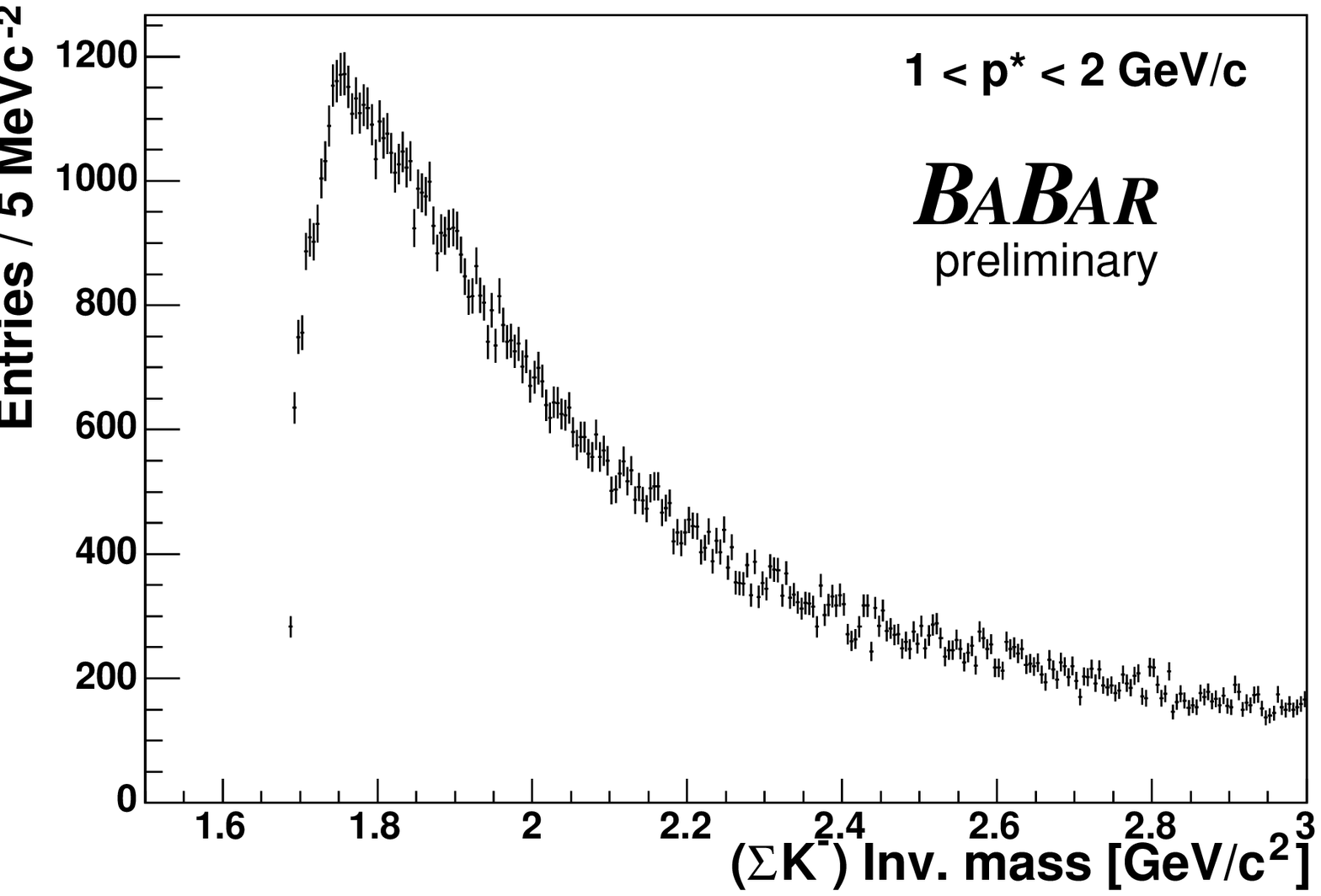}
\vspace*{-0.05truecm}

\includegraphics[height=5.4cm]{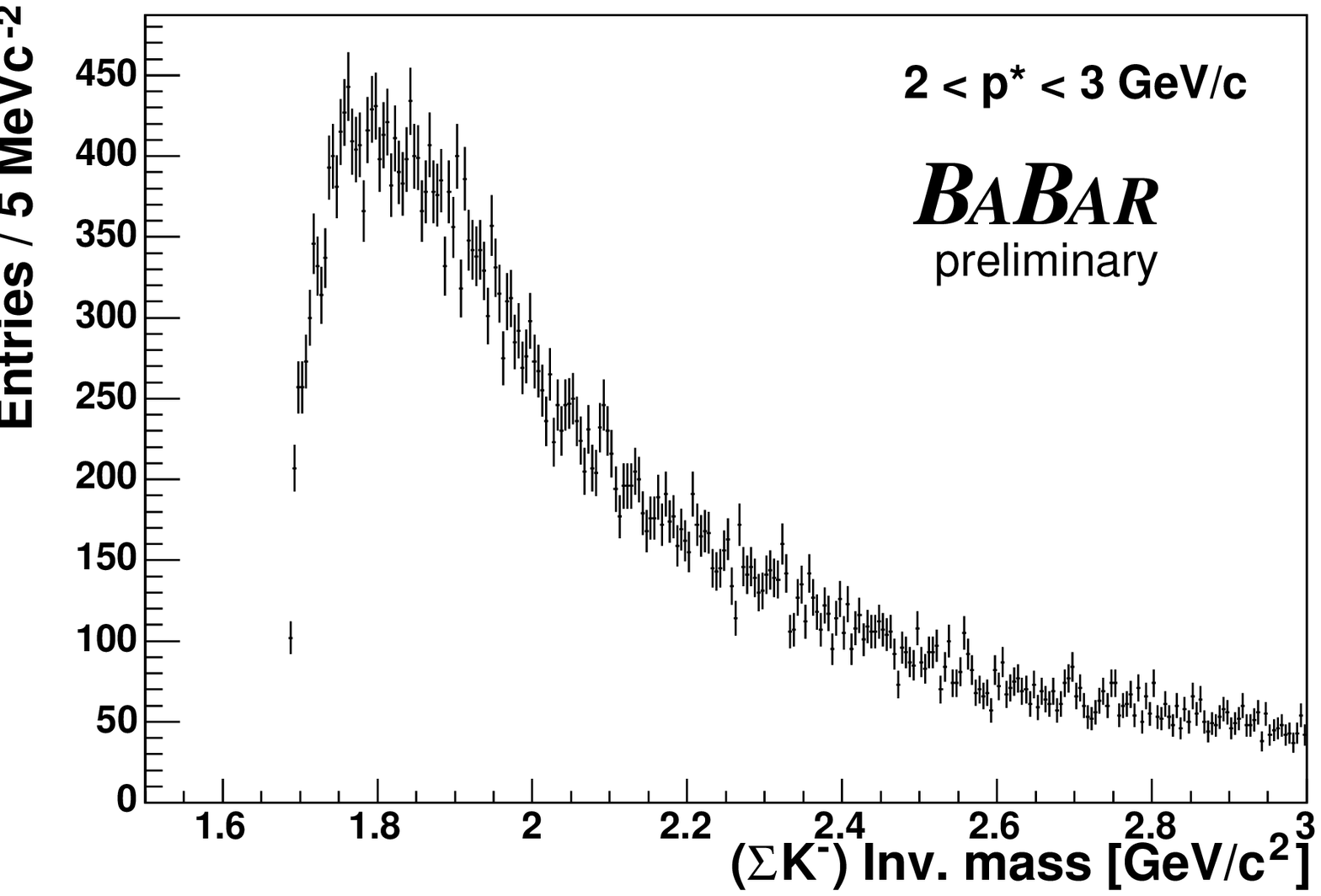}
\includegraphics[height=5.4cm]{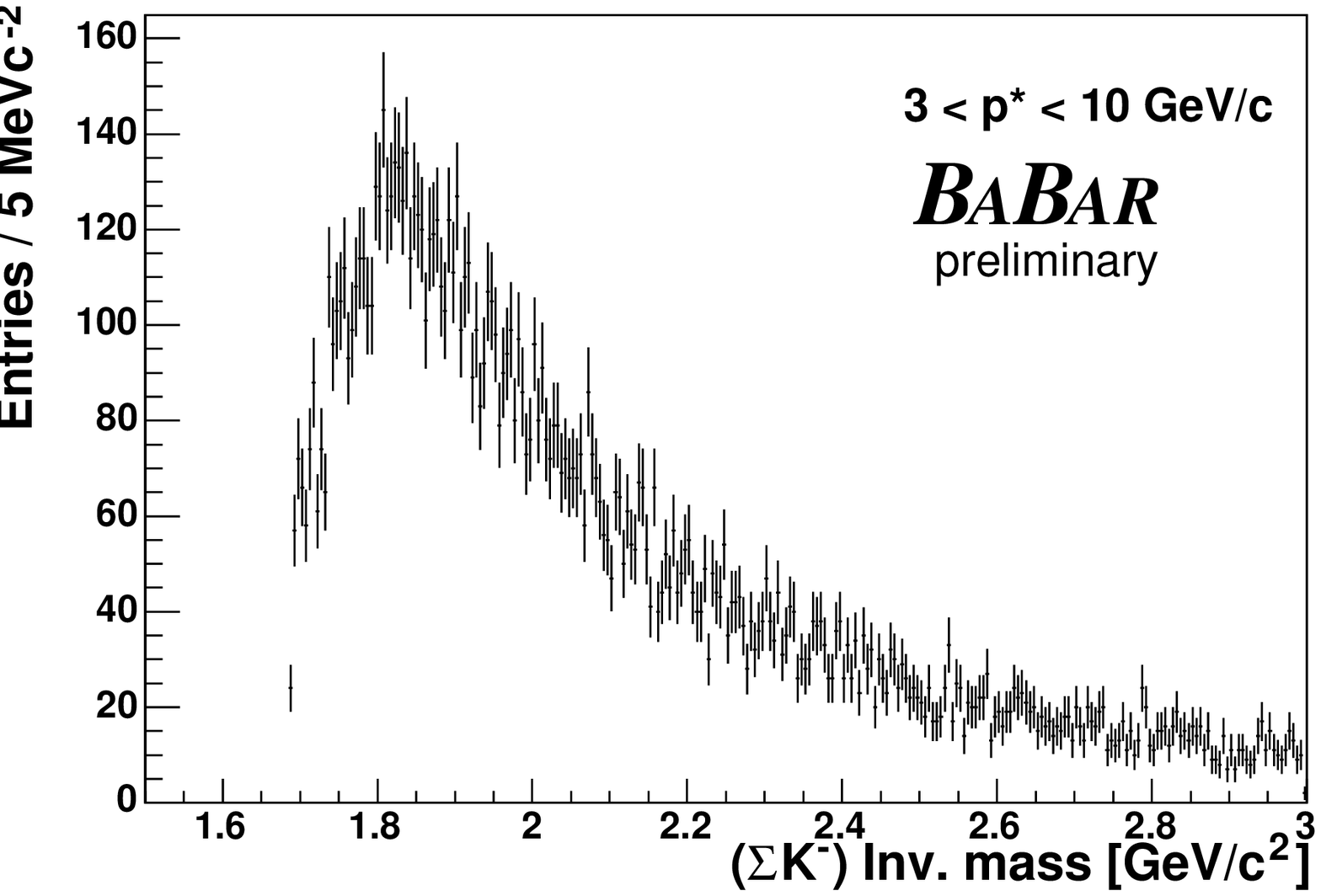}
\vspace{-0.5truecm}
\caption{
Invariant mass distributions of $\SigzKm$ for $p^*$ in four different ranges.}
\label{fig:msigkm}
\end{center}  
\end{figure}

\begin{figure}[hbt]
\begin{center}
\vspace*{-0.9truecm}
\scalebox{0.6}{
\includegraphics[height=10.1cm]{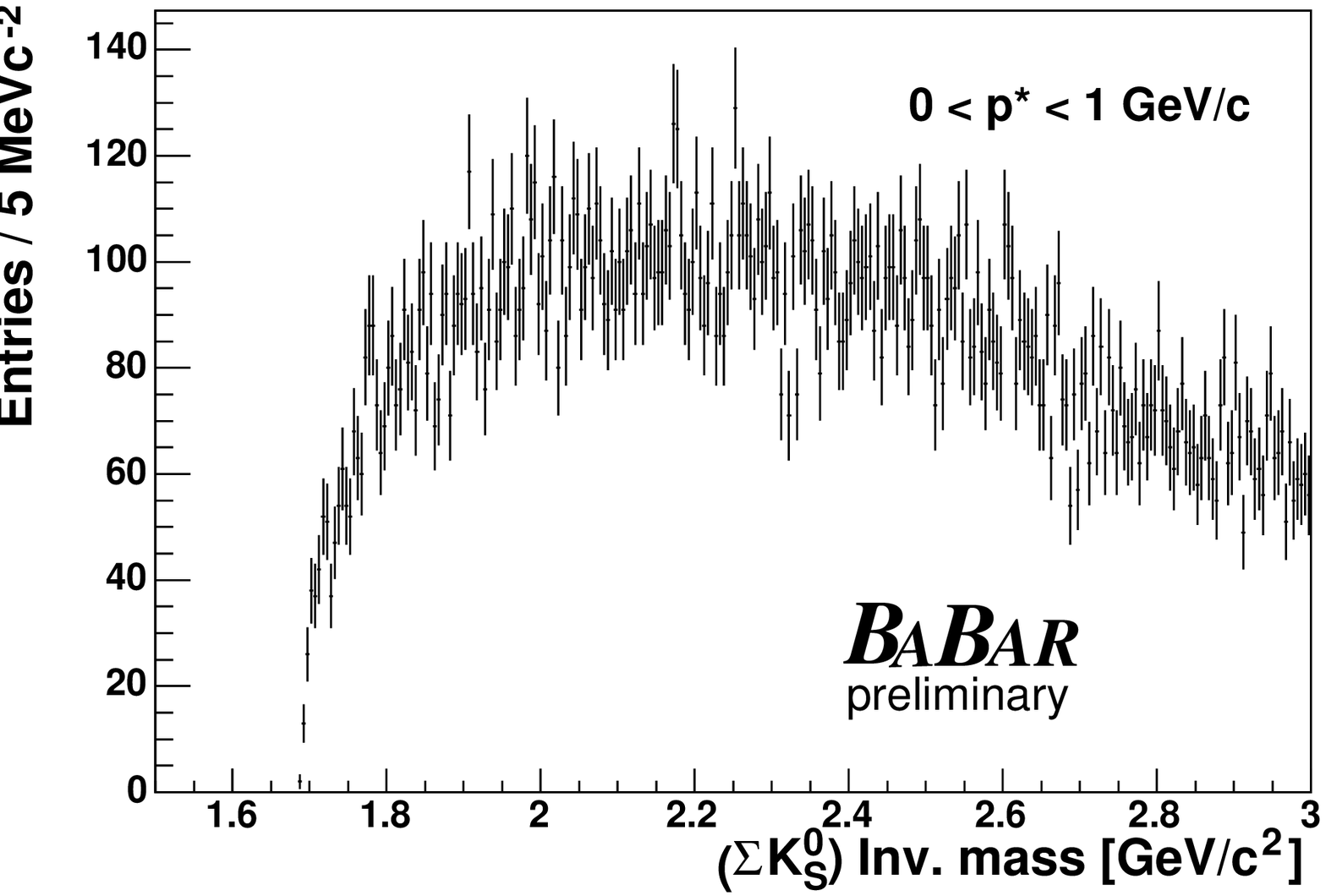}
\includegraphics[height=10.1cm]{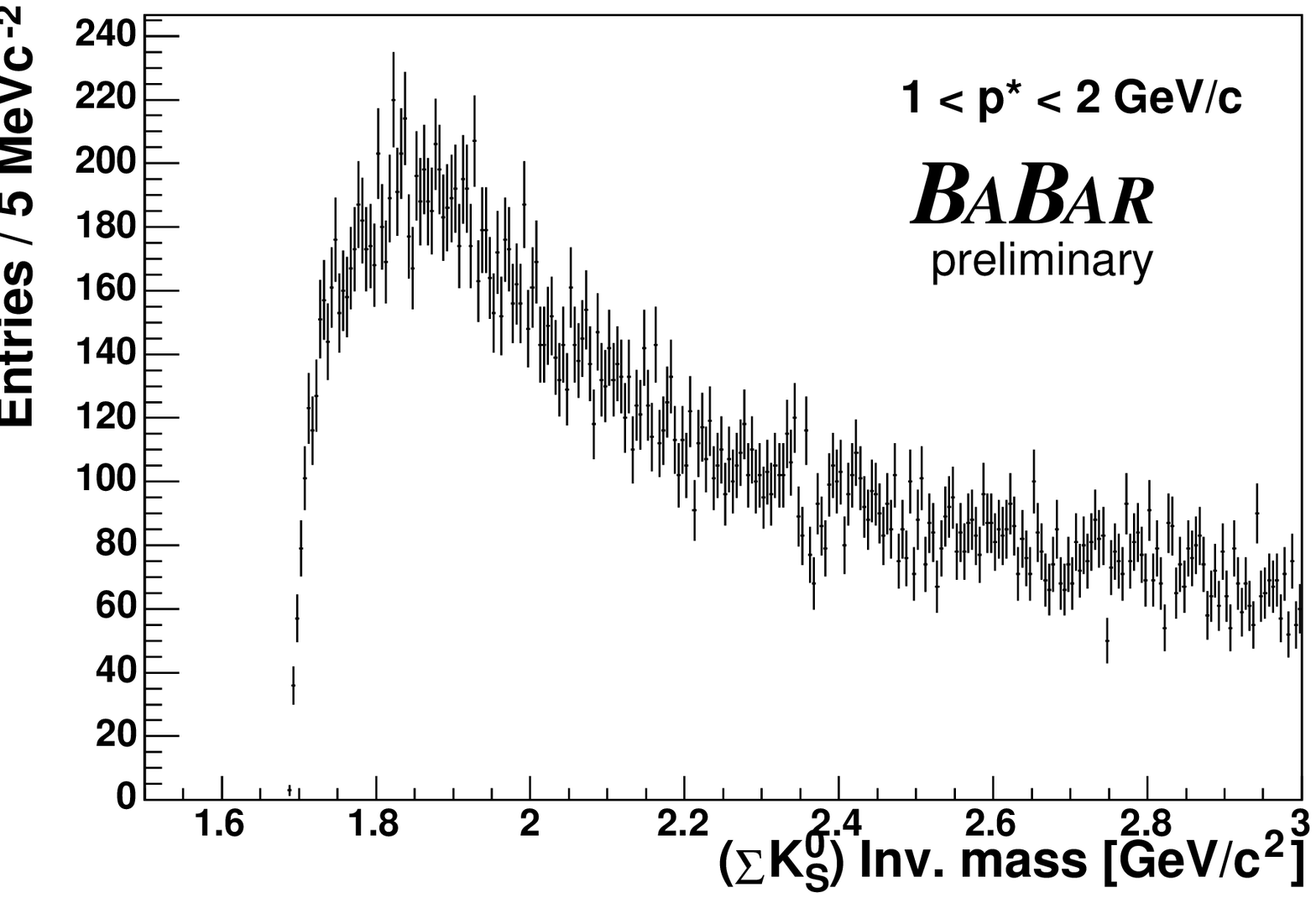}}
\vspace*{-0.5truecm}

\scalebox{0.6}{
\includegraphics[height=10.1cm]{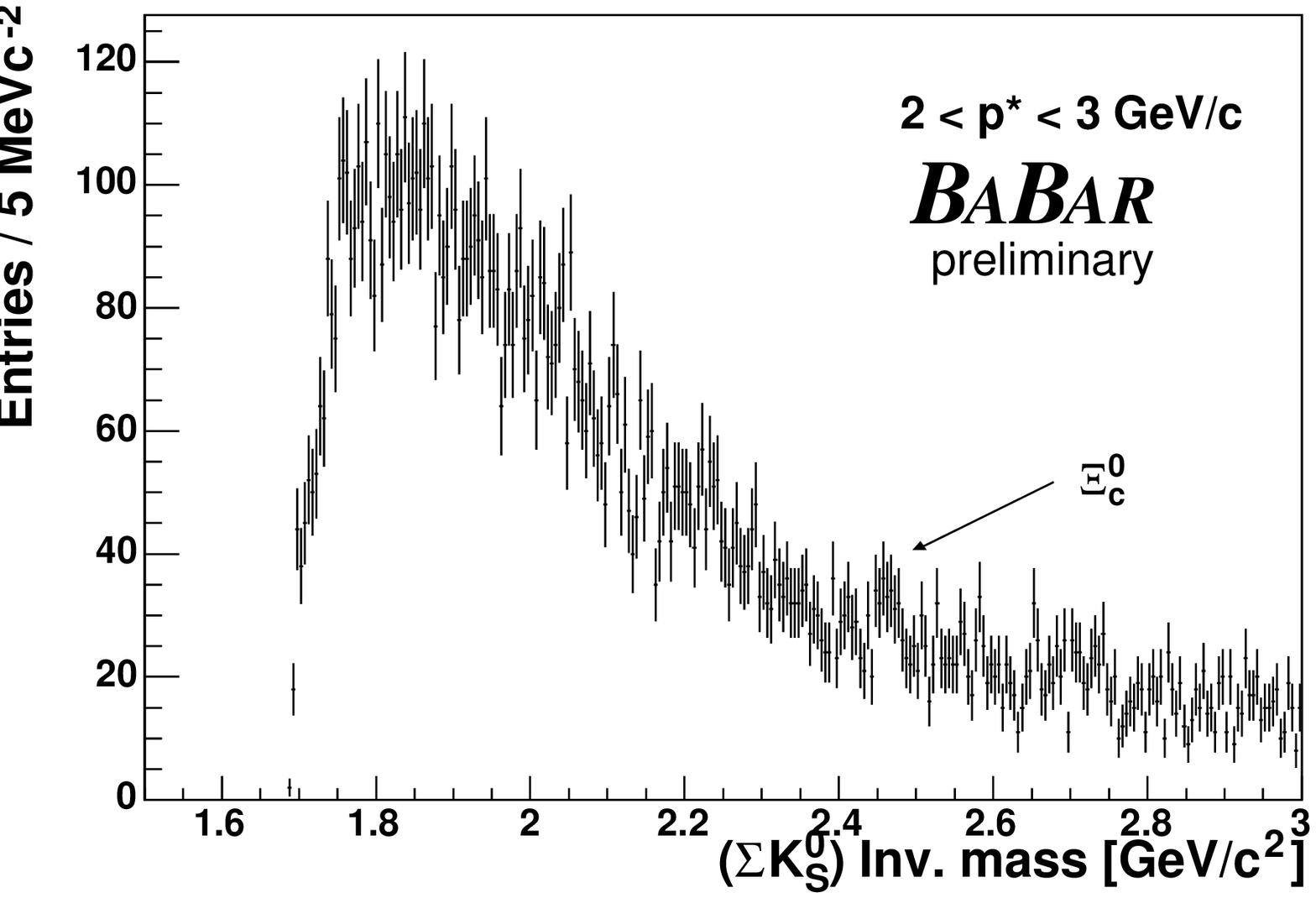}
\includegraphics[height=10.1cm]{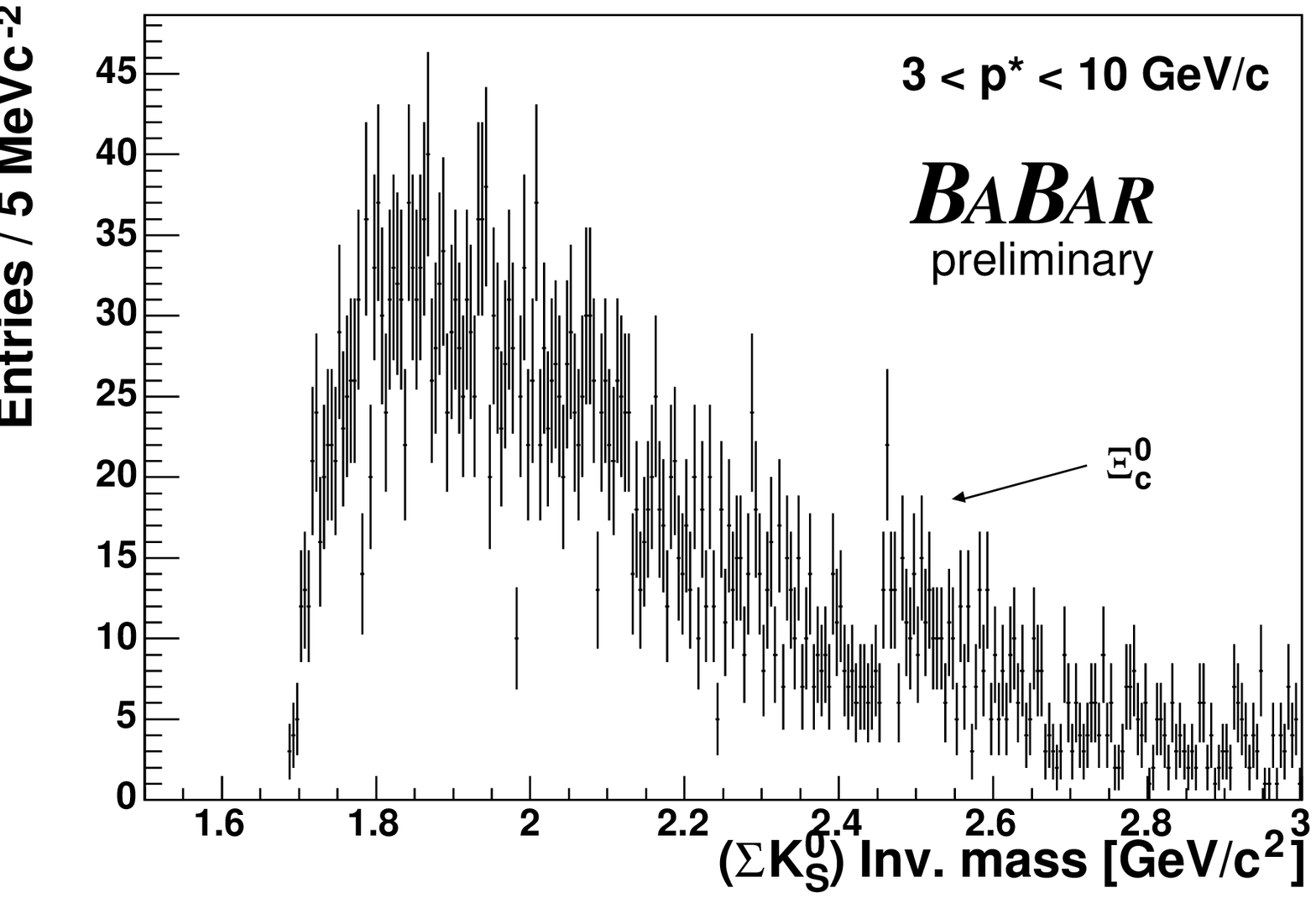}}
\vspace*{-0.8truecm}
\caption{
Invariant mass distributions of $\SigzKs$ for $p^*$ in four different ranges.} 
\label{fig:msigkz}
\end{center}  
\end{figure}

We find a substantial \pstar dependence in the structure of the
invariant mass distributions, and so we show them in four \pstar bins, 
for $\Lambda^0 K^+$, $\Lambda^0 K^-$ and $\Lambda^0 \KS$ combinations in
Figs.~\ref{fig:mlamkp}, \ref{fig:mlamkm} and \ref{fig:mlamkz}, respectively,
and for $\Sigma^0 K^+$, $\Sigma^0 K^-$ and $\Sigma^0 \KS$ combinations
in Figs.~\ref{fig:msigkp}, \ref{fig:msigkm} and \ref{fig:msigkz}, respectively.
In Fig.~\ref{fig:mlamkp} there is a peak as expected from
$\Lambda_c^+$(2285) and also some structure in the 1950--2150 \mevcc region
that can be attributed to $\Lambda_c^+$ decays with one or two
missing pions.
No other narrow peaks are evident.

In Fig.~\ref{fig:mlamkm} the \Omeg peak is prominent and
there is a hint of a broad structure corresponding to the
$\Xi^-(1820)$,
but no other narrow structure is visible.
In Fig.~\ref{fig:mlamkz} we observe the $\Xi_c^0$(2472) at high
momenta, and there are hints of the $\Xi^0(1820)$, 
but no other narrow structure.
The $\Sigma^0K$ modes have much lower statistics, but show structure
consistent with the corresponding $\Lambda^0K$ mode.
There is a hint of the $\Xi^-(1950)$ at low \pstar in the $\Sigma^0K^-$ mode.
In no case is any unexpected narrow structure seen, 
and all evident structure is consistent with that due to known resonances.
We have also examined the data in finer \pstar bins and find
no sign of a pentaquark signal in any \pstar range.

We perform fits in ten \pstar bins to the $\Lambda^0K^-$ and
$\Lambda^0\KS$ mass distributions, in which we might expect signals
near 1860 \mevcc from the \xmpq and \xmn states, respectively.
The resolution functions derived from the 
data and simulation are described by sums of two
Gaussians with total rms of about 4 \mevcc that depend slightly on \pstar.
For each Breit-Wigner width we consider both 1 and 18 \mevcc, 
as for the $\xmm$.
The fits are performed over the mass range 
from 1772 \mevcc to 1972 \mevcc
in order to exclude known resonances.
The background function is a first-order polynomial.

\begin{figure}[hbt]
\begin{center}
\scalebox{0.6}{
\includegraphics[width=13.7cm]{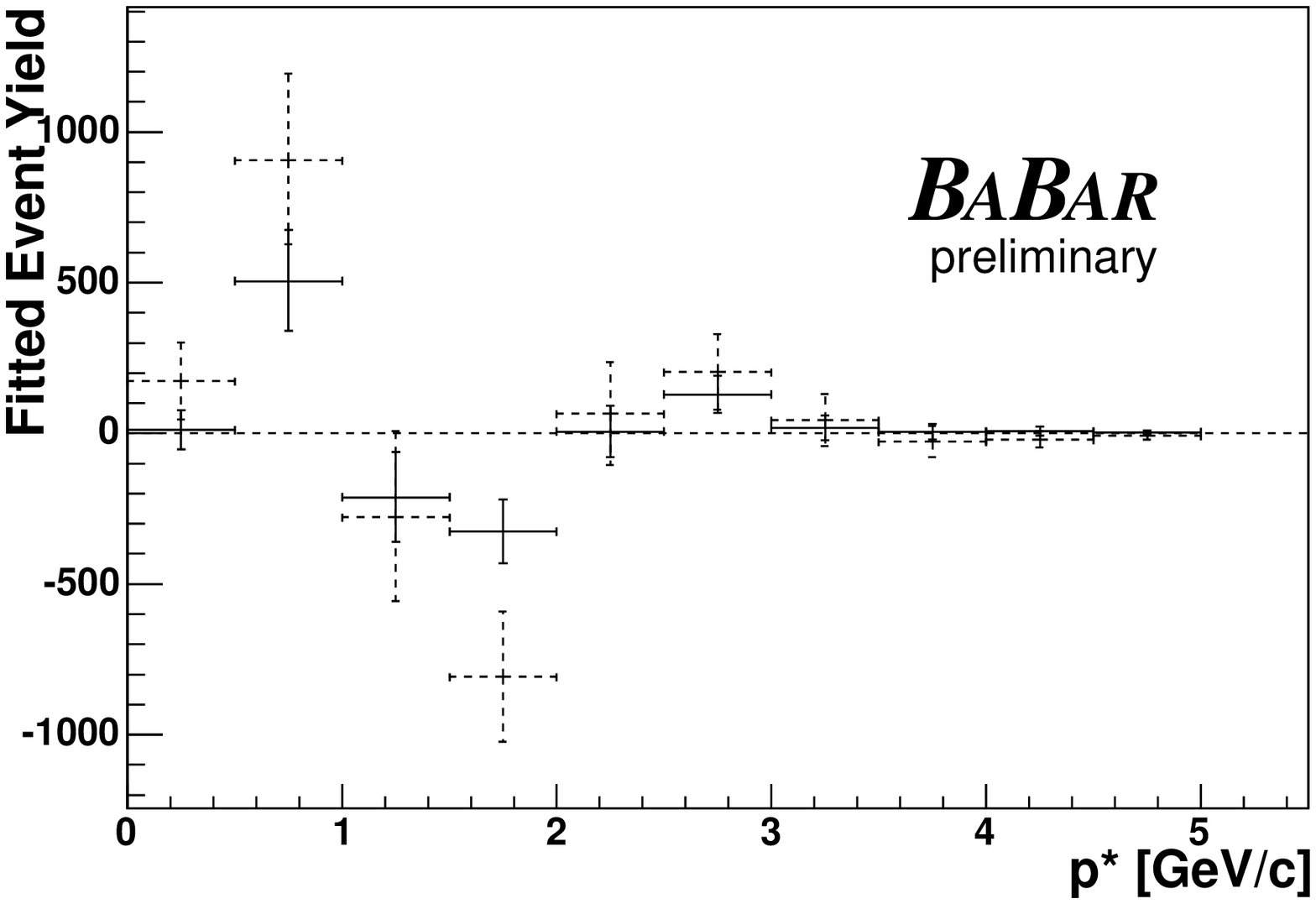}
\includegraphics[width=13.7cm]{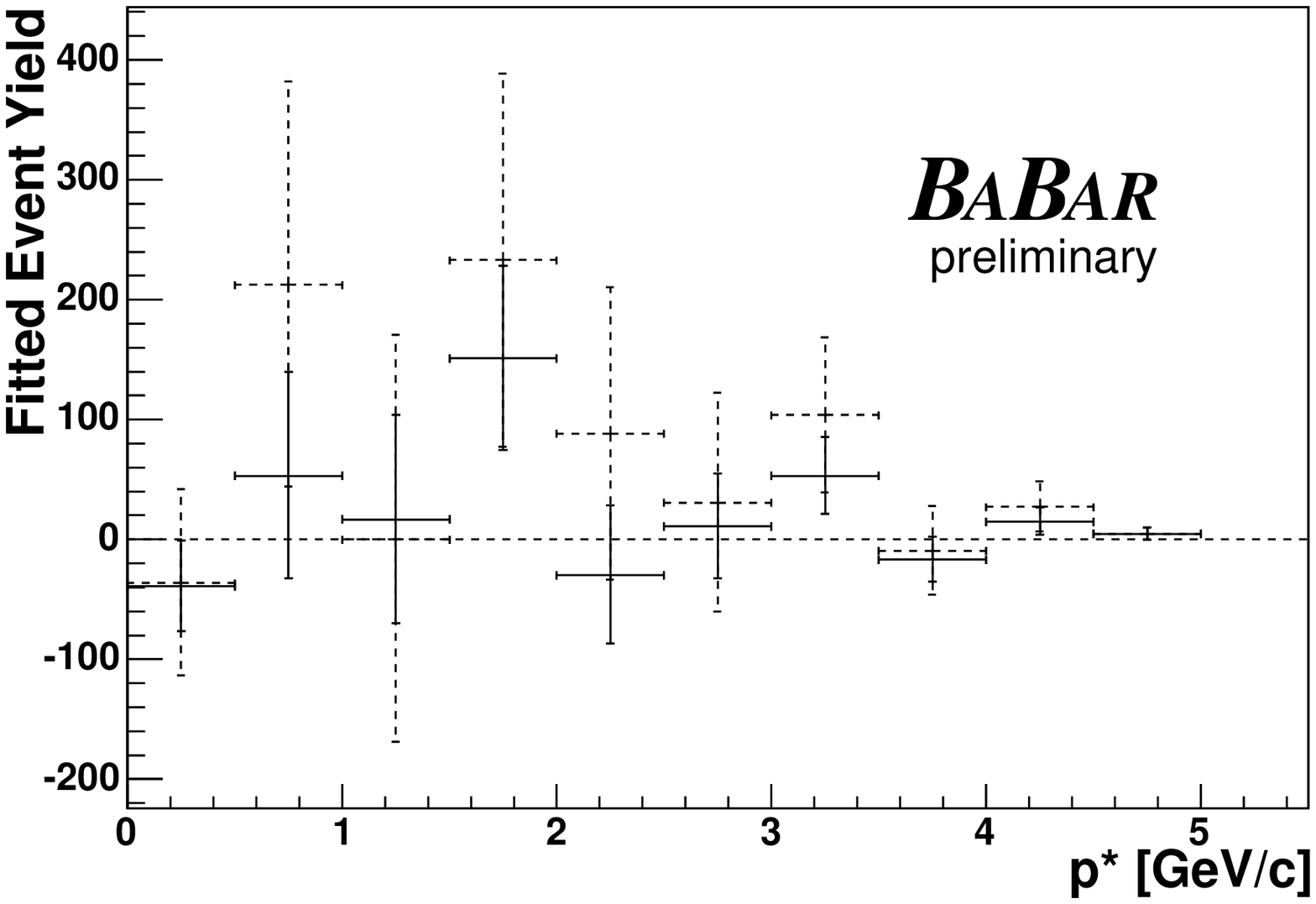}}
\end{center}
\vspace{-0.8truecm}
\caption {The \xmpq (left) and \xmn (right) signal yields extracted
  from the fits to the $\Lambda^0K^-$ and $\Lambda^0\KS$ mass
  distributions, respectively, assuming a mass of 1862 \mevcc and 
  width  $\Gamma=1$ (solid) and  $\Gamma=18$ \mevcc (dashed).
} 
\label{fig:xnsignal}
\end{figure}

We fix the masses to 1862 \mevcc, plus a shift obtained from the
simulation in each \pstar bin that does not exceed 2 \mevcc,
and obtain the signal yields shown in Fig.~\ref{fig:xnsignal}.
In all \pstar bins the fit quality is good across the full mass range and the 
signal is consistent with zero.
Systematic uncertainties on the fitting procedure are again found to be
negligible compared with the statistical uncertainties, and variations
of the mass and selection criteria give consistent results.

\clearpage

\section{Upper Limits}
\label{sec:results}

For the states reported by previous experiments, there exist specific masses
at which to search and experimental upper limits on the decay widths.
We can therefore calculate upper limits on their differential 
production cross sections, with some assumptions on the branching fraction 
to the mode in which the search was made.
We present such limits for the \Th, \xmm, \xmpq and \xmn.

For the $\Th(1540) \to p\KS$ we take the yields from the fits shown
above (see Fig.~\ref{fig:thppq}) and convert them into cross
sections by dividing by the efficiency (including the $\KS\to\pim\pip$
branching fraction), the integrated luminosity and the \pstar bin width.
The reconstruction efficiency is calculated from the simulation and
corrected using data, and varies smoothly from 13\% at low \pstar to
22\% at high \pstar.
It is checked by measuring the differential cross section for 
$\Lambda_c^+ \to p\KS$ in the combination of $q\bar{q}$ and \Y4S events 
represented in our data,
which is found to be consistent with the appropriate combination of 
previous measurements.
If the \Th is a $udud\bar{s}$ pentaquark state, 
we expect only two possible decay modes, $nK^+$ and $pK^0$,
with very similar $Q$ values and hence roughly equal branching fractions.
Assuming half of the $K^0$ appear as $K^0_S$, we arrive at a branching
fraction $B(\Th \to p\KS)=1/4$, and we divide by this value to obtain
the total \Th differential production cross section.

We then derive a conservative upper limit on this cross section in
each bin.
We consider only the physically allowed region, and scale the limit
by a factor of $(1+\delta\epsilon/\epsilon)$, where
$\delta\epsilon/\epsilon = 0.049$ is the sum in quadrature of the
relative systematic uncertainties on the efficiency and luminosity.
The relative luminosity uncertainty is 1\%, and that on the efficiency
is dominated by the 3.2\% uncertainty on the reconstruction of pairs
of tracks from a displaced vertex such as a \KS.
The total is nearly independent of \pstar.

\begin{figure}[hbt]
\begin{center}
\scalebox{0.6}{
\includegraphics[width=13.9cm]{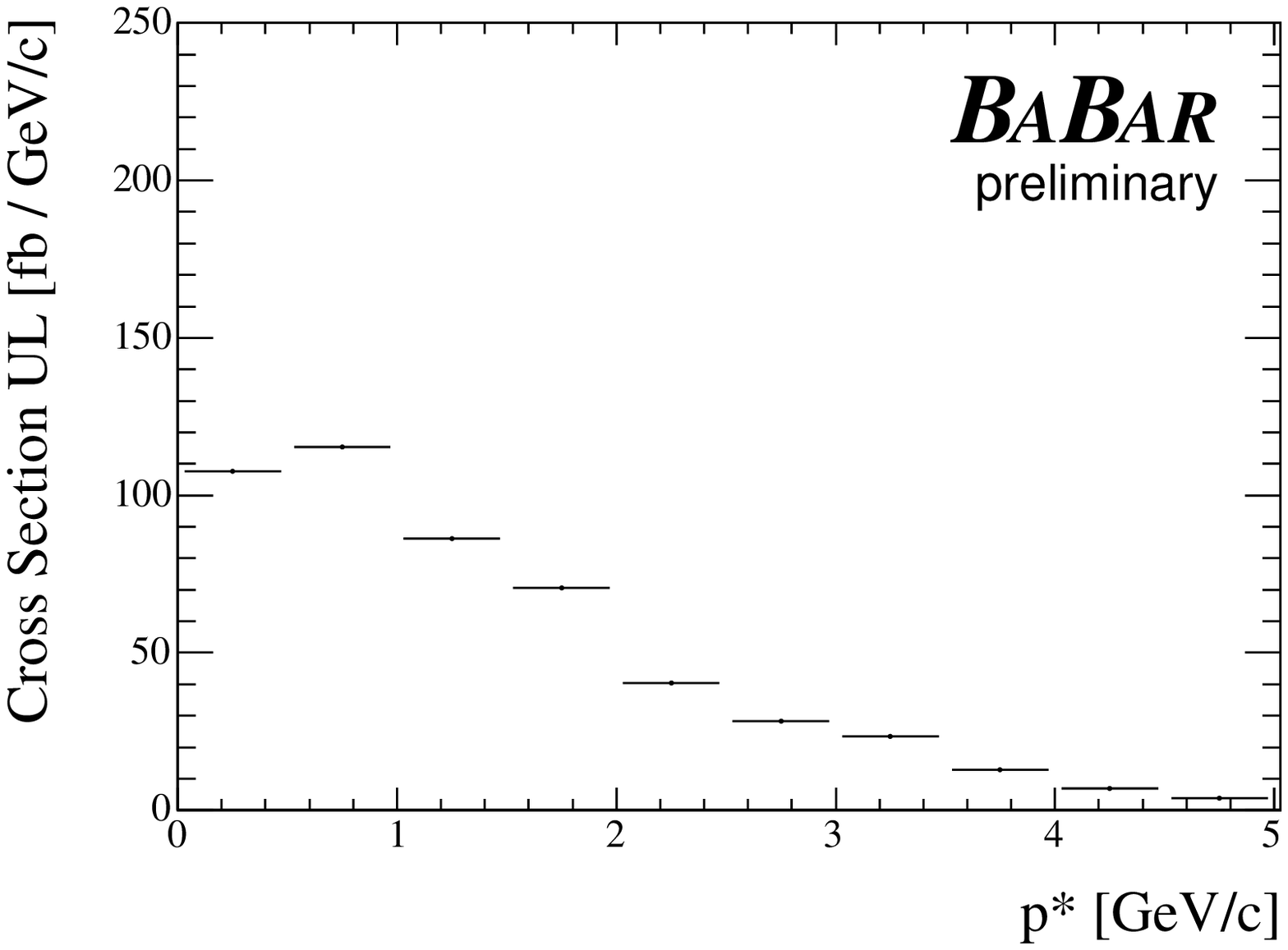}
\includegraphics[width=13.9cm]{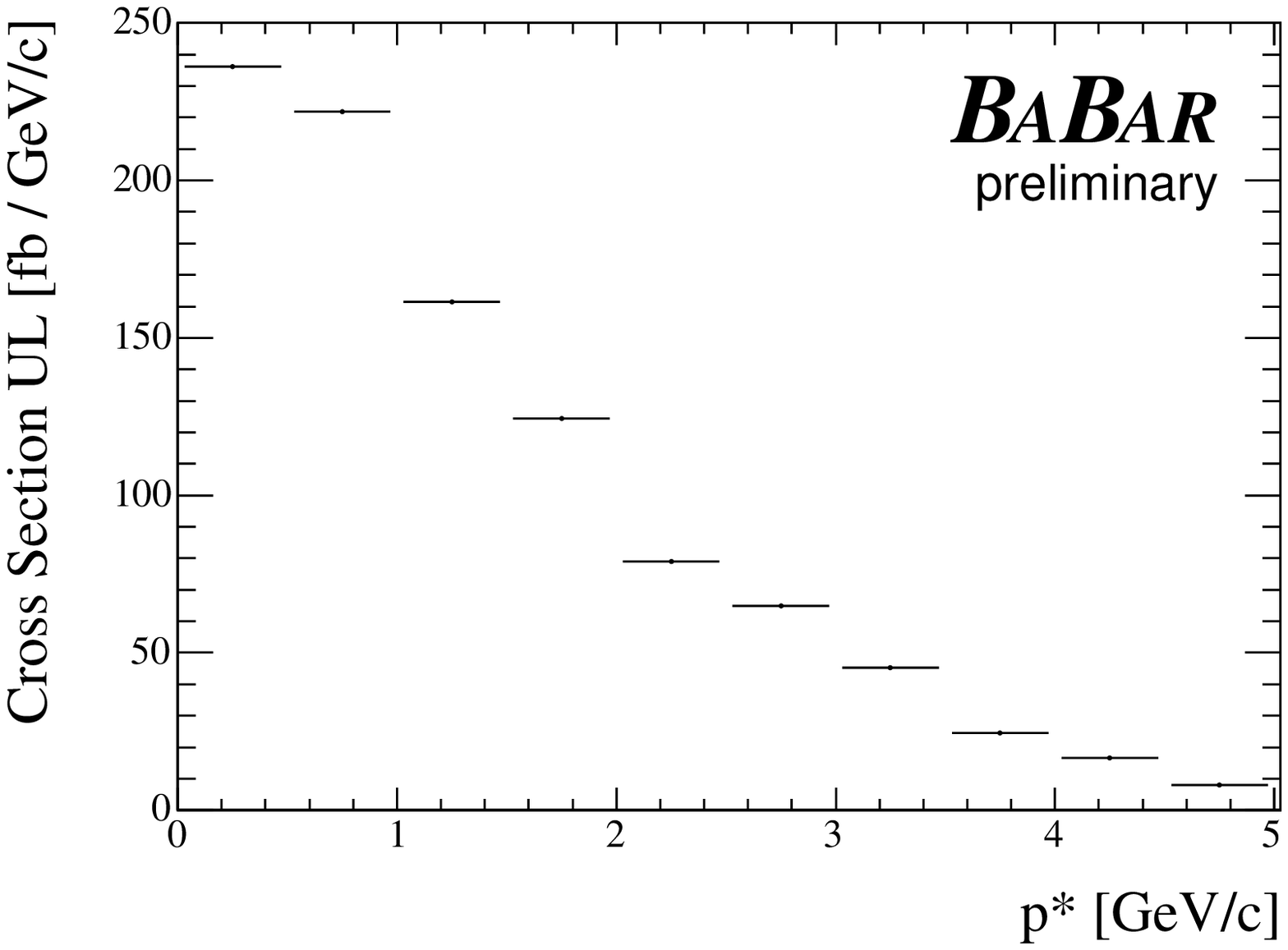}}
\end{center}
\vspace{-0.8truecm}
\caption {
The $95\%$ C.L. upper limit on the production cross-section for the \Th 
assuming a mass of 1540 \mevcc and a natural width $\Gamma=1$ \mevcc
(left) or $\Gamma=8$ \mevcc (right), as a function of c.m.\ momentum \pstar.}
\label{fig:firstul}
\end{figure}

These upper limits are shown in Fig.~\ref{fig:firstul}
and tabulated in Table~\ref{tab:ulth}.
Two sets of limits are shown; one corresponds to a very narrow state, 
and the other to a width at the current experimental upper limit of 8 \mevcc.
The limits correspond to the mass of 1540 \mevcc used in the fits;
repeating the analysis at several nearby mass values gives consistent
limits.
The units are fb per \gevc , 
and apply to the sum of all possible production processes.
To isolate continuum and \Y4S\ production we divide by the respective 
cross sections, and the corresponding limits on the numbers of
pentaquarks per event are given in 
Table~\ref{tab:ulth}.

Similarly we convert the measured yields described above for the
$\xmm (1862)\to \xm\pim$ (Fig.~\ref{fig:xmsignal}),
$\xmpq(1862)\to \Lambda^0K^-$ (Fig.~\ref{fig:xnsignal}) and
$\xmn (1862)\to \Lambda^0\KS$ (Fig.~\ref{fig:xnsignal}) decays 
into cross sections.
Here we assume a mass of 1862 \mevcc and present limits for both
a very narrow hypothesis and for $\Gamma=18$ \mevcc.
The $\xmm (1862)\to \xm\pim$ reconstruction efficiency varies smoothly 
from 6.5\% at low \pstar to 12\% at high \pstar, and has been checked
using the observed $\Xi^{*0}(1530)\to \xm\pip$ signal.
Its relative systematic uncertainty again varies slightly with \pstar,
and its average value of $\delta\epsilon/\epsilon = 0.064$
is larger than for the $p\KS$ mode, as there are two displaced
vertices and more tracks in the decay.
The $\xmpq(1862)\to \Lambda^0K^-$ efficiency varies
from 10\% at low \pstar to 27\% at high \pstar, is checked with the 
$\Omega^-\to \Lambda^0K^-$ signal, and the average 
$\delta\epsilon/\epsilon = 0.052$ is similar to the $p\KS$ mode.
The $\xmn(1862)\to \Lambda^0\KS$ efficiency varies
from 3.5\% at low \pstar to 13\% at high \pstar, 
and $\delta\epsilon/\epsilon = 0.072$
is similar to the $\xm\pim$ mode.

\begin{figure}[hbt]
\begin{center}
\scalebox{0.6}{
\includegraphics[width=16.0cm]{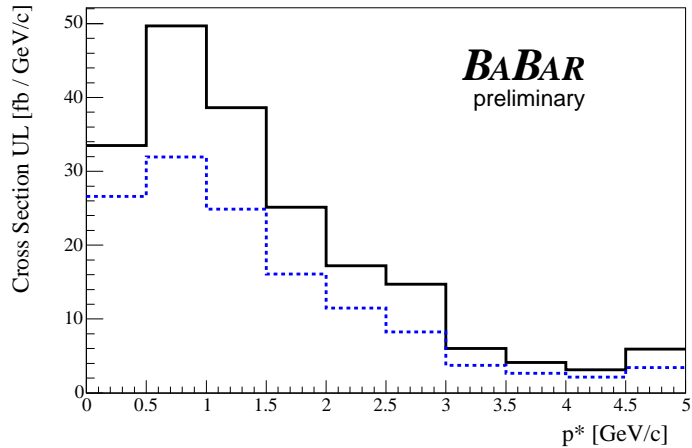}}
\end{center}
\vspace{-0.8truecm}
\caption {
The $95\%$ C.L. upper limit on the production cross-section for \xmm
assuming a natural width of
$\Gamma=1$ (dashed) and $\Gamma=18$ \mevcc (solid), as a function of
c.m.\ momentum \pstar.}
\label{fig:secondul}
\end{figure}

\begin{figure}[hbt]
\begin{center}
\scalebox{0.6}{
\includegraphics[width=13.7cm]{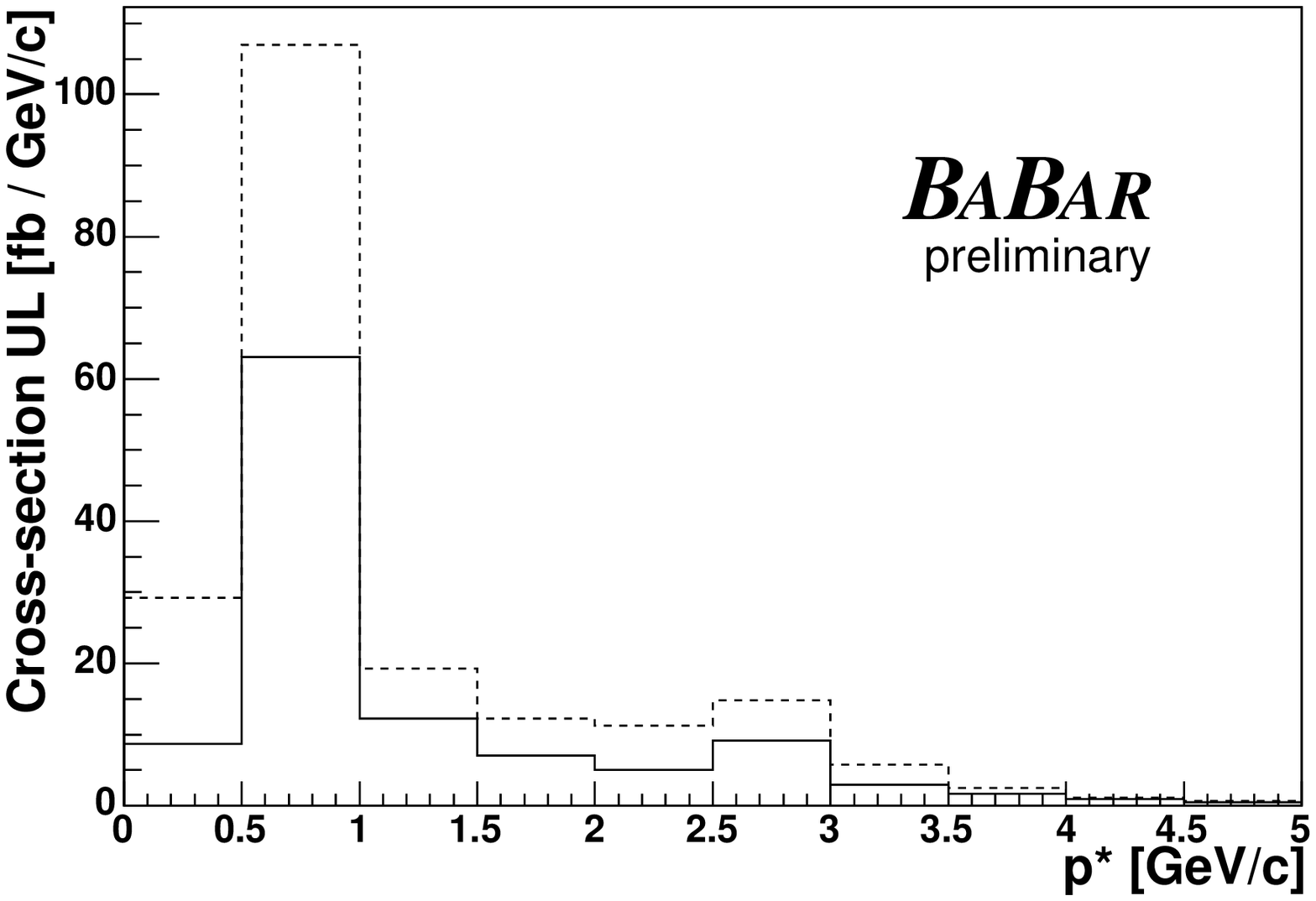}
\includegraphics[width=13.7cm]{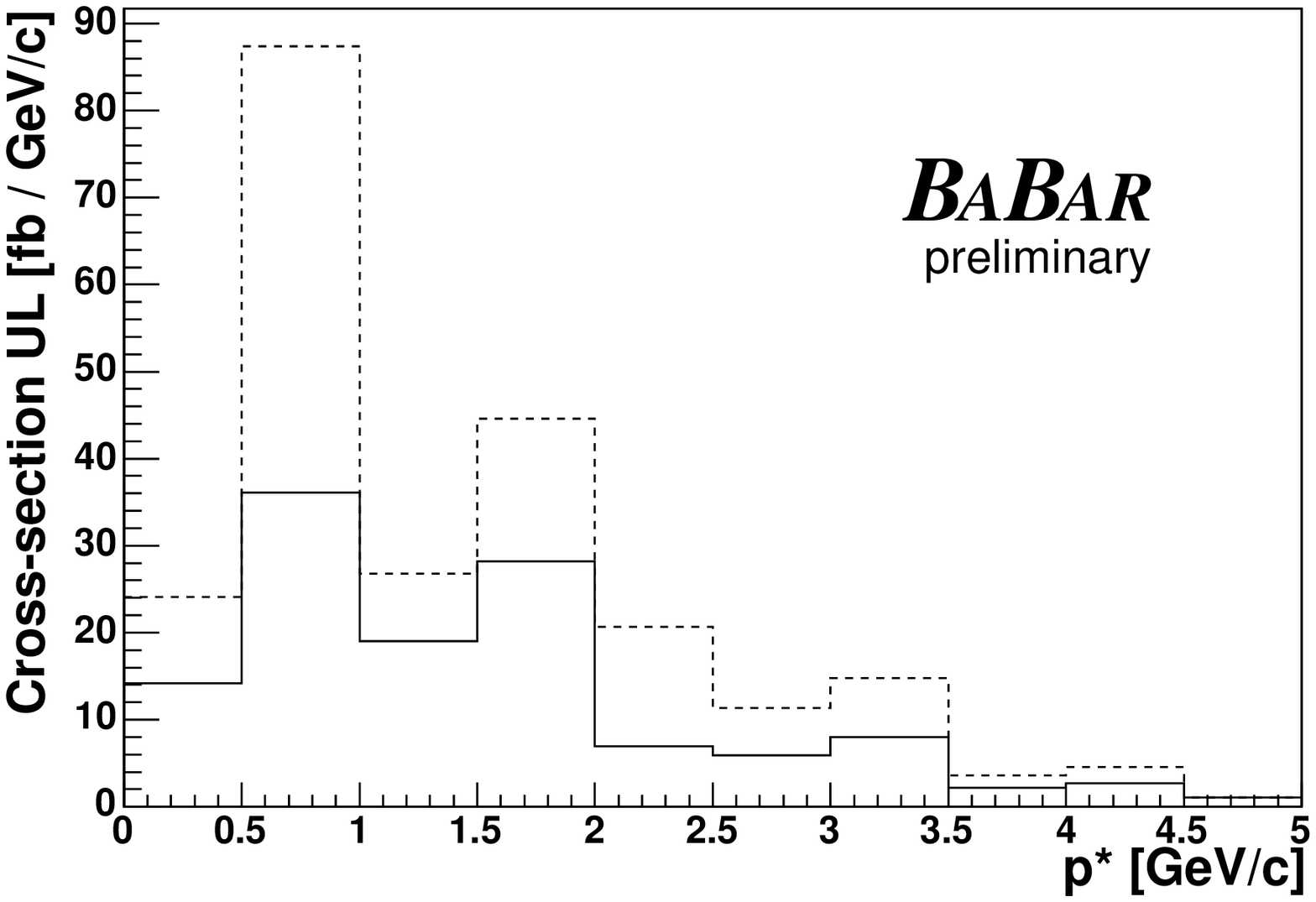}}
\end{center}
\vspace{-0.8truecm}
\caption {
The $95\%$ C.L. upper limits on the production cross-section for \xmpq
times its branching fraction into $\Lambda^0K^-$ (left), and \xmn times
its branching fraction into $\Lambda^0\KS$ (right), assuming a natural width of
$\Gamma=1$ \mevcc (solid) and $\Gamma=18$ \mevcc (dashed), 
as a function of c.m.\ momentum \pstar.}
\label{fig:lastul}
\end{figure}

We assume a $\xmm \to \xm\pim$ branching fraction of one-half; 
for the \xmpq and \xmn we do not assume branching fractions into the measured
modes,
but present limits on the product of the production cross section and
branching fraction.
Again the results are independent of the assumed mass; they are
shown in Figs.~\ref{fig:secondul} and~\ref{fig:lastul}, and tabulated in 
Tables~\ref{tab:ulxmm}, \ref{tab:ulxmq} and~\ref{tab:ulxmn}.

\begin{table}[hbt]
\begin{center} 
\caption{
The measured \Th signal yield in each \pstar bin, assuming a mass of
1540 \mevcc and two values of the natural width $\Gamma$ in \mevcc.
The corresponding upper limits at the 95\% C.L.\ on
the \Th differential production cross section,
the number of \Th produced per $e^+e^- \to q\bar{q}$ event and
the number per \Y4S decay.
}
\vspace{0.2cm}
\begin{small}
\begin{tabular}{|c|r@{$\pm$}r|r@{$\pm$}r||r|r||r|r||r|r|} \hline 
             \multicolumn{11}{|c| }{$\Th$(1540) (preliminary)}\\ \hline
\pstar Range & \multicolumn{4}{|c||}{Yield}
             & \multicolumn{2}{|c||}{X-section U.L.}
             & \multicolumn{2}{|c||}{$q\bar{q}$ event U.L.}
             & \multicolumn{2}{|c| }{\Y4S decay U.L.} \\ [.1cm]
 (\gevc)     & \multicolumn{4}{|c||}{($p\KS$ mode)}
             & \multicolumn{2}{|c||}{(fb/GeVc$^{-1}$)}
             & \multicolumn{2}{|c||}{($10^{-5}$/(evt$\cdot$GeVc$^{-1}$))}
             & \multicolumn{2}{|c| }{($10^{-5}$/(evt$\cdot$GeVc$^{-1}$))}
\\[.2cm]
             & \multicolumn{2}{|c| }{$\Gamma=1$}
             & \multicolumn{2}{|c||}{$\Gamma=8$} & $\Gamma=1$ & $\Gamma=8$
             & $\Gamma=1$ & $\Gamma=8$& $\Gamma=1$ & $\Gamma=8$  \\ \hline
 & \multicolumn{2}{|c|}{ } & \multicolumn{2}{|c||}{ } &&&&&& \\ [-.3cm]
0.0--0.5     &   107 &$^{139}_{134}$  &   288  &$^{283}_{242}$ 
             &   107.6 & 236.2 & 3.17 & 6.97   &10.76 &23.62 \\[0.1cm]
0.5--1.0     &$-$178 &$^{238}_{236}$  &$-$194  &$^{457}_{452}$  
             &   115.3 & 221.8 & 3.40 &   6.54 &11.53 &22.18 \\[0.1cm]
1.0--1.5     &    19 &$^{210}_{208}$  & $-$31  &$^{406}_{410}$  
             &    86.2 & 161.5 & 2.54 &   4.76 & 8.61 &16.14 \\[0.1cm]
1.5--2.0     &   109 &$^{156}_{156}$  &   112  &$^{307}_{310}$  
             &    70.7 & 124.4 & 2.09 &   3.67 & 7.07 &12.44 \\[0.1cm]
2.0--2.5     &$-$158 &$^{114}_{113}$  &$-$322  &$^{222}_{243}$  
             &    40.4 &  79.0 & 1.19 &   2.33 & 2.83 & 7.90 \\[0.1cm]
2.5--3.0     & $-$38 &$^{83}_{83}$    &    41  &$^{175}_{164}$  
             &    28.3 &  64.8 & 0.83 &   1.91 & 2.35 & 6.48 \\[0.1cm]
3.0--3.5     &    33 &$^{59}_{59}$    &    50  &$^{119}_{117}$  
             &    23.4 &  45.2 & 0.69 &   1.33 &---  & ---   \\[0.1cm]
3.5--4.0     & $-$11 &$^{39}_{39}$    & $-$56  &$^{74}_{80}$    
             &    12.9 &  24.5 & 0.38 &   7.23 &---  & ---   \\[0.1cm]
4.0--4.5     &     4 &$^{22}_{21}$    &    25  &$^{47}_{44}$   
             &     6.8 &  16.6 & 0.20 &   4.88 &---  & ---   \\[0.1cm]
4.5--5.0     &     1 &$^{12}_{12}$    &     6  &$^{24}_{24}$   
             &     3.7 &   7.9 & 0.11 &   2.33 &---  & ---   \\[0.1cm]
\hline\hline
Total        & \multicolumn{2}{|c| }{---}  & \multicolumn{2}{|c||}{---}
             & 182.8 & 363.1 & 5.39 &10.71 &17.90 &34.98 \\
             & \multicolumn{2}{|c|}{  } & \multicolumn{2}{|c||}{  }
             & \multicolumn{2}{|c||}{  }& \multicolumn{2}{|c||}{  }
             & \multicolumn{2}{|c|}{  } \\[-.3cm]
             & \multicolumn{2}{|c|}{  } & \multicolumn{2}{|c||}{  }
             & \multicolumn{2}{|c||}{(fb)}
             & \multicolumn{2}{|c||}{($10^{-5}$/event)}
             & \multicolumn{2}{|c|}{($10^{-5}$/event)} \\
\hline
\end{tabular}
\end{small}
\label{tab:ulth}
\end{center}
\end{table}

\begin{table}[hbt]
\begin{center} 
\caption{
The measured \xmm signal yield in each \pstar bin, assuming a mass of
1862 \mevcc and two values of the natural width in \mevcc.
The corresponding upper limits at the 95\% C.L.\ on
the \xmm differential production cross section,
the number of \xmm produced per $e^+e^- \to q\bar{q}$ event and
the number per \Y4S decay.}
\vspace{0.2cm}
\begin{small}
\begin{tabular}{|c|r|r||r|r||r|r||r|r|} \hline 
             \multicolumn{9}{|c| }{$\xmm \to \xm\pim$ (preliminary)}\\  \hline
\pstar Range & \multicolumn{2}{|c||}{Yield}
           & \multicolumn{2}{|c||}{X-section U.L.}
           & \multicolumn{2}{|c||}{$q\bar{q}$ event U.L.}
             & \multicolumn{2}{|c| }{\Y4S decay U.L.} \\ [.1cm]
 (\gevc)     & \multicolumn{2}{|c||}{  }
             & \multicolumn{2}{|c||}{(fb/GeVc$^{-1}$)}
             & \multicolumn{2}{|c||}{($10^{-5}$/(evt$\cdot$GeVc$^{-1}$))}
             & \multicolumn{2}{|c| }{($10^{-5}$/(evt$\cdot$GeVc$^{-1}$))} 
\\[.2cm]
           & $\Gamma=1$  & $\Gamma=18$  & $\Gamma=1$ & $\Gamma=18$
           & $\Gamma=1$  & $\Gamma=18$  & $\Gamma=1$ & $\Gamma=18$  \\ \hline
0.0--0.5   &   13$\pm$26 &  $-$7$\pm$37 & 26.6 &33.5 & 0.78& 0.99 & 2.53&3.19\\
0.5--1.0   &$-$62$\pm$52 &$-$128$\pm$67 & 31.9 &49.7 & 0.94& 1.46 & 3.04&4.73\\
1.0--1.5   &$-$56$\pm$50 & $-$63$\pm$76 & 24.9 &38.6 & 0.73& 1.14 & 2.37&3.68\\
1.5--2.0   &$-$70$\pm$52 &$-$102$\pm$62 & 16.1 &25.1 & 0.47& 0.74 & 1.53&2.39\\
2.0--2.5   &$-$48$\pm$33 & $-$64$\pm$50 & 11.5 &17.2 & 0.34& 0.51 & 1.09&1.64\\
2.5--3.0   &    5$\pm$28 &    21$\pm$35 &  8.3 &14.7 & 0.24& 0.43 & 0.79&1.40\\
3.0--3.5   &$-$25$\pm$15 & $-$36$\pm$28 &  3.7 & 6.0 & 0.11& 0.18 &---  & ---\\
3.5--4.0   &    5$\pm$10 &     8$\pm$12 &  2.6 & 4.1 & 0.08& 0.12 &---  & ---\\
4.0--4.5   &    3$\pm$10 &     2$\pm$11 &  2.1 & 3.1 & 0.06& 0.09 &---  & ---\\
4.5--5.0   &    1$\pm$12 &     1$\pm$22 &  3.4 & 5.9 & 0.10& 0.17 &---  & ---\\
\hline\hline
Total      &     ---     &     ---      & 22.0 &33.7 & 0.65& 0.99 & 2.11&3.20\\
             & \multicolumn{2}{|c||}{  }
             & \multicolumn{2}{|c||}{  }& \multicolumn{2}{|c||}{  }
             & \multicolumn{2}{|c|}{  } \\[-.3cm]
             & \multicolumn{2}{|c||}{  }
             & \multicolumn{2}{|c||}{(fb)}
             & \multicolumn{2}{|c||}{($10^{-5}$/event)}
             & \multicolumn{2}{|c|}{($10^{-5}$/event)} \\
\hline
\end{tabular}
\end{small}
\label{tab:ulxmm}
\end{center}
\end{table}

\begin{table}[hbt]
\begin{center} 
\caption{
The measured \xmpq signal yield in each \pstar bin, assuming a mass of
1862 \mevcc and 
two values of the natural width in \mevcc.
The corresponding upper limits at the 95\% C.L.\ on
the \xmpq differential production cross section times its branching fraction 
$B$ into $\Lambda^0K^-$,
$B\times$the number of \xmpq produced per $e^+e^- \to q\bar{q}$ event and
$B\times$the number per \Y4S decay.}
\vspace{-.2cm}
\begin{small}
\begin{tabular}{|c|r@{$\pm$}r|r@{$\pm$}r||r|r||r|r||r|r|} \hline 
        \multicolumn{11}{|c| }{$\xmpq \to \Lambda^0K^-$ (preliminary)}\\ hline
\pstar Range & \multicolumn{4}{|c||}{Yield}
             & \multicolumn{2}{|c||}{X-section U.L.}
             & \multicolumn{2}{|c||}{$q\bar{q}$ event U.L.}
             & \multicolumn{2}{|c| }{\Y4S decay U.L.} \\ [.1cm]
 (\gevc)     & \multicolumn{4}{|c||}{   }
             & \multicolumn{2}{|c||}{(fb/ GeVc$^{-1}$)}
             & \multicolumn{2}{|c||}{($10^{-5}$/(evt$\cdot$GeVc$^{-1}$))}
             & \multicolumn{2}{|c| }{($10^{-5}$/(evt$\cdot$GeVc$^{-1}$))} 
\\[.2cm]
             & \multicolumn{2}{|c| }{$\Gamma=1$}
             & \multicolumn{2}{|c||}{$\Gamma=18$} & $\Gamma=1$ & $\Gamma=18$
             & $\Gamma=1$ & $\Gamma=18$& $\Gamma=1$ & $\Gamma=18$  \\ \hline
 & \multicolumn{2}{|c|}{ } & \multicolumn{2}{|c||}{ } &&&&&& \\ [-.3cm]
0.0--0.5 &    11 &$^{ 65}_{ 64}$ &    174 &$^{128}_{127}$ & 40.9 & 122.4 & 1.21 & 3.61 & 3.90 &11.66 \\[0.06cm]
0.5--1.0 &   505 &$^{171}_{164}$ &    906 &$^{287}_{277}$ &220.9 & 385.7 & 6.52 &11.38 &21.04 &36.73 \\[0.06cm]
1.0--1.5 &$-$213 &$^{151}_{148}$ & $-$278 &$^{285}_{278}$ & 62.8 & 118.8 & 1.85 & 3.50 & 5.98 &11.31 \\[0.06cm]
1.5--2.0 &$-$326 &$^{106}_{105}$ & $-$808 &$^{216}_{215}$ & 34.3 &  69.8 & 1.01 & 2.06 & 3.27 & 6.65 \\[0.06cm]
2.0--2.5 &     6 &$^{ 85}_{ 84}$ &     65 &$^{172}_{170}$ & 24.7 &  57.2 & 0.73 & 1.69 & 2.35 & 5.45 \\[0.06cm]
2.5--3.0 &   129 &$^{ 62}_{ 61}$ &    203 &$^{126}_{125}$ & 33.0 &  59.5 & 0.97 & 1.76 &  --- &  --- \\[0.06cm]
3.0--3.5 &    17 &$^{ 42}_{ 41}$ &     43 &$^{ 87}_{ 86}$ & 12.8 &  27.5 & 0.38 & 0.81 &  --- &  --- \\[0.06cm]
3.5--4.0 &     4 &$^{ 26}_{ 26}$ &  $-$28 &$^{ 53}_{ 52}$ &  7.5 &  13.8 & 0.22 & 0.41 &  --- &  --- \\[0.06cm]
4.0--4.5 &     7 &$^{ 15}_{ 14}$ &  $-$21 &$^{ 28}_{ 27}$ &  3.9 &   6.0 & 0.12 & 0.18 &  --- &  --- \\[0.06cm]
4.5--5.0 &     3 &$^{  6}_{  5}$ &  $-$ 8 &$^{ 13}_{ 12}$ &  1.9 &   3.3 & 0.06 & 0.10 &  --- &  --- \\[0.06cm]
\hline\hline
Total        & \multicolumn{2}{|c| }{---}  & \multicolumn{2}{|c||}{---}
                                                          & 83.6 & 181.0 & 2.76 & 5.34 & 7.87 &15.88 \\
             & \multicolumn{2}{|c|}{  } & \multicolumn{2}{|c||}{  }
             & \multicolumn{2}{|c||}{  }& \multicolumn{2}{|c||}{  }
             & \multicolumn{2}{|c|}{  } \\[-.35cm]
             & \multicolumn{2}{|c|}{  } & \multicolumn{2}{|c||}{  }
             & \multicolumn{2}{|c||}{(fb)}
             & \multicolumn{2}{|c||}{($10^{-5}$/event)}
             & \multicolumn{2}{|c|}{($10^{-5}$/event)} \\
\hline
\end{tabular}
\end{small}
\label{tab:ulxmq}
\end{center}
\end{table}

\begin{table}[hbt]
\begin{center} 
\caption{
The measured \xmn signal yield in each \pstar bin, assuming a mass of
1862 \mevcc and 
two values of the natural width in \mevcc.
The corresponding upper limits at the 95\% C.L.\ on
the \xmn differential production cross section times its branching fraction 
$B$ into $\Lambda^0\KS$,
$B\times$the number of \xmn produced per $e^+e^- \to q\bar{q}$ event and
$B\times$the number per \Y4S decay.}
\vspace{0.2cm}
\begin{small}
\begin{tabular}{|c|r@{$\pm$}r|r@{$\pm$}r||r|r||r|r||r|r|} \hline 
             \multicolumn{11}{|c| }{$\xmn \to \Lambda^0\KS$ (preliminary)}\\ \hline
\pstar Range & \multicolumn{4}{|c||}{Yield}
             & \multicolumn{2}{|c||}{X-section U.L.}
             & \multicolumn{2}{|c||}{$q\bar{q}$ event U.L.}
             & \multicolumn{2}{|c| }{\Y4S decay U.L.} \\ [.1cm]
 (\gevc)     & \multicolumn{4}{|c||}{   }
             & \multicolumn{2}{|c||}{(fb/GeVc$^{-1}$)}
             & \multicolumn{2}{|c||}{($10^{-5}$/(evt$\cdot$GeVc$^{-1}$))}
             & \multicolumn{2}{|c| }{($10^{-5}$/(evt$\cdot$GeVc$^{-1}$))}
\\[.2cm]
             & \multicolumn{2}{|c| }{$\Gamma=1$}
             & \multicolumn{2}{|c||}{$\Gamma=18$} & $\Gamma=1$ & $\Gamma=18$
             & $\Gamma=1$ & $\Gamma=18$& $\Gamma=1$ & $\Gamma=18$  \\ \hline
 & \multicolumn{2}{|c|}{ } & \multicolumn{2}{|c||}{ } &&&&&& \\ [-.3cm]
0.0--0.5 & $-$39 &$^{ 38}_{ 38}$ &$-$36 &$^{ 78}_{ 77}$ &  46.7 &  95.8 & 1.38 & 2.83 &  4.45 &  9.12 \\[0.06cm]
0.5--1.0 &    53 &$^{ 87}_{ 85}$ &  212 &$^{169}_{168}$ & 105.7 & 256.4 & 3.12 & 7.56 & 10.07 & 24.42 \\[0.06cm]
1.0--1.5 &    17 &$^{ 87}_{ 87}$ &    0 &$^{171}_{169}$ &  60.6 & 108.5 & 1.79 & 3.20 &  5.77 & 10.34 \\[0.06cm]
1.5--2.0 &   151 &$^{ 78}_{ 76}$ &  233 &$^{155}_{156}$ &  71.0 & 126.2 & 2.09 & 3.72 &  6.76 & 12.02 \\[0.06cm]
2.0--2.5 &   -30 &$^{ 58}_{ 57}$ &   88 &$^{122}_{122}$ &  23.0 &  65.3 & 0.68 & 1.93 &  2.19 &  6.22 \\[0.06cm]
2.5--3.0 &    11 &$^{ 44}_{ 43}$ &   31 &$^{ 92}_{ 91}$ &  18.1 &  39.1 & 0.54 & 1.15 &  ---  &   --- \\[0.06cm]
3.0--3.5 &    53 &$^{ 32}_{ 32}$ &  104 &$^{ 65}_{ 64}$ &  20.3 &  40.2 & 0.60 & 1.19 &  ---  &   --- \\[0.06cm]
3.5--4.0 & $-$17 &$^{ 19}_{ 19}$ & $-$9 &$^{ 37}_{ 36}$ &   6.8 &  13.2 & 0.20 & 0.39 &  ---  &   --- \\[0.06cm]
4.0--4.5 &    15 &$^{ 12}_{ 11}$ &   27 &$^{ 21}_{ 21}$ &   6.9 &  12.6 & 0.20 & 0.37 &  ---  &   --- \\[0.06cm]
4.5--5.0 &     5 &$^{  6}_{  5}$ &    5 &$^{  6}_{  5}$ &   2.7 &   2.7 & 0.08 & 0.08 &  ---  &   --- \\[0.06cm]
\hline \hline
Total        & \multicolumn{2}{|c| }{---}  & \multicolumn{2}{|c||}{---}
                                                        &   82.8&   204.7 & 2.44 & 6.04 &  7.25 & 18.02 \\
             & \multicolumn{2}{|c|}{  } & \multicolumn{2}{|c||}{  }
             & \multicolumn{2}{|c||}{  }& \multicolumn{2}{|c||}{  }
             & \multicolumn{2}{|c|}{  } \\[-.35cm]
             & \multicolumn{2}{|c|}{  } & \multicolumn{2}{|c||}{  }
             & \multicolumn{2}{|c||}{(fb)}
             & \multicolumn{2}{|c||}{($10^{-5}$/event)}
             & \multicolumn{2}{|c|}{($10^{-5}$/event)} \\
\hline
\end{tabular}
\end{small}
\label{tab:ulxmn}
\end{center}
\end{table}

\large

In order to quote limits on the total production cross section (times
branching fraction) we must either 
know the momentum spectrum or believe that it does not vary rapidly on the 
scale of our bin size.
Since the latter is true in the simulation, we assign conservative, 
model-independent upper limits on the total numbers of pentaquarks
produced per
$e^+e^-\to q\bar{q}$ event (\Y4S\ decay) by summing each
differential cross section over the \pstar range from zero to the
kinematic limit for 5.3 \gev jets ($B$ meson decays), 
taking into account the fact that most of the systematic error is
common to all bins.
Limits are derived from these sums by the same method as in each bin,
and are listed in Tables~\ref{tab:ulth}--\ref{tab:ulxmn}.
Any postulated spectrum can be folded with our differential limit to obtain a 
limit on the total cross section assuming that spectrum, which will
be lower than the model independent limits given in the tables.

\section{Summary}
\label{sec:summary}

We have performed a preliminary high statistics search for the reported states
\Th(1540), \xmm(1862), \xmpq(1862) and \xmn(1862) in $e^+e^-$ annihilations, 
and also for most of the other members of the pentaquark octet 
and anti-decuplet to which they are postulated to belong.
In all cases we observe clear signals for known baryon
resonances that demonstrate sensitivity to any new narrow resonances,
with invariant mass resolution better than the reported upper limits
on the widths of the respective states.
We find no evidence for the production of such states 
in \ldata of \babar\ data.
For the reported states, 
we see no excess at the measured invariant mass values 
and use the reported limits on their widths to set upper limits on the
their inclusive production in our search modes
in both $e^+e^- \to q\bar{q}$ events events and \Y4S\ decays.
These limits are at the level of $10^{-4}$--$10^{-5}$ per event,
depending on the width assumed, and are valid for any narrow state in
the vicinity of 1540 or 1860 \mevcc.
The searches for other members of the multiplets show no evidence for
an unknown narrow resonance at any mass between threshold for the
decay mode used and the kinematic limit.

\begin{figure}[hbt]
\begin{center}
\scalebox{0.6}{
\includegraphics[width=26.9cm]{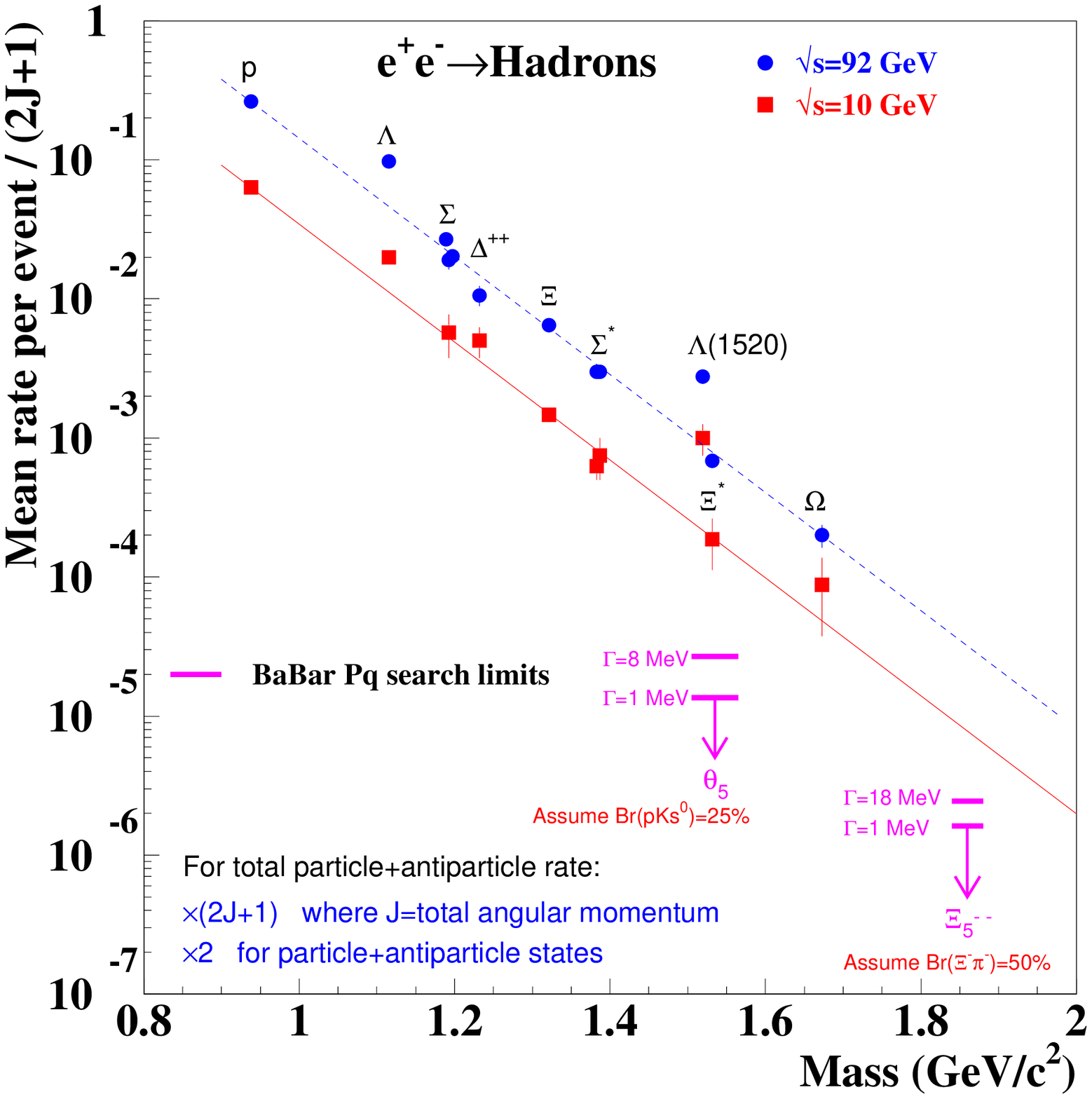}}
\end{center}
\caption {
Compilation of baryon production rates in $e^+e^-$ 
annihilation~\cite{ref:pdg2002}
from experiments at the $Z^0$ (circles) and $\sqrt{s}\approx 10$ \gev
(squares) as a function of baryon mass.  
The vertical scale accounts for the number of spin and particle$+$antiparticle 
states, and the lines are chosen to guide the eye.  
The arrows indicate our preliminary upper limits on spin-1/2 \Th and
\xmm pentaquark 
states, assuming the branching fractions shown, and are seen to lie below the 
solid line.
}
\label{fig:bprod}
\end{figure}

In order to limit the production rates we must know the
branching fraction of each state into the decay mode used in the search.
In the case of the \Th, we take this to be 1/4, as noted above.
For \xmm a similar argument that $\xm \pim$ and $\Sigma^- K^-$ dominate and 
have roughly equal branching fractions leads to a
branching fraction of 1/2.
Taking the upper limit widths,
we calculate 95\% C.L. upper limits on the total production rates of
$1.1\times10^{-4}$ \Th and $1.0\times10^{-5}$ \xmm per 
$e^+e^- \to q\bar{q}$ event (preliminary);
these are roughly a factor of eight and four below the typical values measured 
for ordinary octet and decuplet baryons of the same masses of $8\times10^{-4}$
and $4\times10^{-5}$, respectively, as illustrated in Fig.~\ref{fig:bprod}.

The situation is more complex for the \xmpq and \xmn as the mixing 
between the members of the anti-decuplet and the octet is unknown.
In these two cases the branching fraction of an antidecuplet state of
mass $~\sim$1860 \mevcc to the measured mode could be anywhere from
zero to $\sim$1/3.
The best limit that could be set, assuming the value of 1/3, would
therefore be roughly $15\times10^{-5}$ \xmpq or \xmn per 
$e^+e^- \to q\bar{q}$ event, which is well above the ``expected" value
of $4\times10^{-5}$.
The study of additional modes is needed to elucidate the production
properties of these states.

\section*{Acknowledgements}
We are grateful for the 
extraordinary contributions of our \pep2\ colleagues in
achieving the excellent luminosity and machine conditions
that have made this work possible.
The success of this project also relies critically on the 
expertise and dedication of the computing organizations that 
support \babar.
The collaborating institutions wish to thank 
SLAC for its support and the kind hospitality extended to them. 
This work is supported by the
US Department of Energy
and National Science Foundation, the
Natural Sciences and Engineering Research Council (Canada),
Institute of High Energy Physics (China), the
Commissariat \`a l'Energie Atomique and
Institut National de Physique Nucl\'eaire et de Physique des Particules
(France), the
Bundesministerium f\"ur Bildung und Forschung and
Deutsche Forschungsgemeinschaft
(Germany), the
Istituto Nazionale di Fisica Nucleare (Italy),
the Foundation for Fundamental Research on Matter (The Netherlands),
the Research Council of Norway, the
Ministry of Science and Technology of the Russian Federation, and the
Particle Physics and Astronomy Research Council (United Kingdom). 
Individuals have received support from 
CONACyT (Mexico),
the A. P. Sloan Foundation, 
the Research Corporation,
and the Alexander von Humboldt Foundation.



\clearpage

\begin{thebibliography}{99} 

\bibitem{prl91012003}
LEPS Collaboration, T.~Nakano \etal, 
Phys.\ Rev.\ Lett. {\bf 91},~012002~(2003).

\bibitem{plb572:127}
SAPHIR Collaboration, J.~Barth \etal,
\newblock Phys.\ Lett.{} {\bf B~572},~127~(2003).

\bibitem{prl92:032001}
CLAS Collaboration, V.~Kubarovsky \etal,
\newblock Phys.\ Rev.\ Lett.{} {\bf 92},~032001~(2004).
\newblock Erratum; ibid, 049902.

\bibitem{diana}
DIANA Collaboration, V.V.~Barmin \etal,
\newblock Phys.\ Atom.\ Nucl.{} {\bf 66},~1715~(2003).

\bibitem{svd}
SVD Collaboration, A.~Aleev \etal,
\newblock hep-ex/0401024~(2004).

\bibitem{hermes}
HERMES Collaboration, A.~Airapetian \etal,
\newblock Phys.\ Lett.{} {\bf B~585},~213~(2004).

\bibitem{lasttheta}
COSY-TOF Collaboration, M.~Abdel-Bary \etal,
\newblock Phys.\ Lett.{} {\bf B~595},~127~(2004).

\bibitem{prl92:042003}
NA49 Collaboration, C. Alt \etal,
\newblock Phys.\ Rev.\ Lett.{} {\bf 92},~042003~(2004).

\bibitem{hep-ex-0403017}
H1 Collaboration, A. Aktas \etal,
\newblock \mbox{hep-ex/0403017} (2004).

\bibitem{review}
See, e.g., J. Pochodzalla, hep-ex/0406077 (2004) and references therein.

\bibitem{zp:a359:305}
D.~Diakonov, V.~Petrov and M.V.~Polyakov,
\newblock Z.\ Phys.{} {\bf A~359},~305~(1997).

\bibitem{Marek}
M.~Karliner and H.~Lipkin,
\newblock hep-ph/0402260 (2004).

\bibitem{Jaffe}
R.~Jaffe and F.~Wilczek,
\newblock Phys.\ Rev.\ Lett. {} {\bf 91},~232003 (2003).

\bibitem{ref:babar}
The \babar\ Collaboration, B.\ Aubert {\em et al.},
Nucl.\ Instrum.\ Methods {\bf A~479}, 1 (2002).

\bibitem{ref:pdg2002}
Particle Data Group, 
K.~Hagiwara {\em et al.}, Phys.\ Rev.\ {\bf D~66}, 010001 (2002).

\bibitem{jetset}
T.~Sjostrand, Comput.\ Phys.\ Commun.\  {\bf 82}, 74 (1994).


\end{thebibliography}
\end{document}